\documentclass[prl,twocolumn,superscriptaddress,aps]{revtex4-1}
\usepackage{amsmath}
\usepackage{amsfonts}
\usepackage{amssymb}
\usepackage{color}
\usepackage{dsfont}

\renewcommand{\dag}{^{\dagger}}
\usepackage{dcolumn}
\usepackage{bm}
\usepackage{slashed}
\usepackage[toc,page]{appendix}
\usepackage{float}
\usepackage[pdftex]{graphicx}
\usepackage[colorlinks=true,citecolor=blue,linkcolor=blue]{hyperref}
\begin{document}
\bibliographystyle{apsrev4-1}

\title{Bulk-Defect Correspondence in Particle-Hole Symmetric Insulators and Semimetals}

\author{Fernando de Juan}
\affiliation{Materials Science Division, Lawrence Berkeley National
Laboratories, Berkeley, CA
94720}
\affiliation{Department of Physics, University of California, Berkeley, CA 94720, USA}
\author{Andreas R\"uegg}
\affiliation{Department of Physics, University of California, Berkeley, CA 94720, USA}
\affiliation{Theoretische Physik, Wolfgang-Pauli-Strasse 27, ETH Z\"urich, CH-8093 Z\"urich,
Switzerland}
\author{Dung-Hai Lee}
\affiliation{Materials Science Division, Lawrence Berkeley National Laboratories, Berkeley, CA
94720}
\affiliation{Department of Physics, University of California, Berkeley, CA 94720, USA}

\date{\today}

\begin{abstract}
Lattices with a basis can host crystallographic defects which share the same topological charge
(e.g.~the Burgers vector $\vec b$ of a dislocation) but differ in their microscopic structure of the
core. We demonstrate that in insulators with particle-hole symmetry and an odd number of orbitals
per site, the microscopic details drastically affect the electronic structure: modifications can
create or annihilate non-trivial bound states with an associated fractional charge. We show that
this observation is related to the behavior of end modes of a dimerized chain and discuss how the
end or defect states are predicted from topological
invariants in these more complicated cases. Furthermore, using explicit examples on the honeycomb
lattice, we explain how bound states in vacancies,
dislocations and disclinations are related to each other and to edge modes and how similar features
arise in nodal semimetals such as graphene.
\end{abstract}

\maketitle

One of the most renowned features of topological insulators (TIs) and superconductors (TSs) is that
they host robust symmetry-protected boundary modes. For example,
quantum spin Hall insulators and 3D TIs have boundary modes protected by time reversal
symmetry $\mathcal{T}$ and charge conservation, while those in TSs are
protected at least by particle-hole symmetry (charge conjugation) $\mathcal{C}$. The general
classification of these phases is well
established \cite{SRFL08,K09,RSF10} and so is the correspondence between bulk invariants and
boundary
modes. 

Another subject that is often discussed in conjunction with topological states is
fermion zero modes in defects \cite{TK10,R10}. For example, in a time reversal invariant insulator
in 3D, whether trivial or not, bound states appear in the core of
a dislocation \cite{RZV09} when the so-called ``weak $\mathbb{Z}_2$ topological invariant" $\vec N$
has the following relation with the Burger's vector $\vec b$
\begin{equation}
\vec N \cdot \vec b = \pi \mod 2\pi.
\label{eq:Ndotb}
\end{equation}
Analogous dislocation states have been studied in other systems including 3D quantum Hall
states \cite{H87}, 2D \cite{JMS12,SMJ13} weak TIs and weak TSs \cite{AN12,UYT13,HYQ13},where they
give rise to Majorana modes \cite{K01}.

In this Letter, we focus on 2D insulators with particle-hole symmetry
($\mathcal{C}$-insulators) only. The subject under consideration is the validity of
Eq.~\eqref{eq:Ndotb} for such systems. The particle-hole symmetry in an insulator, as opposed to a
superconductor, is a fine tuned symmetry that is often absent in a real
system. However, when the symmetry is only weakly broken, the topological analysis is still very
useful in predicting the presence/absence of in-gap defect states \cite{RH02,QHZ08,DUM11}. In the
``ten-fold way" \cite{SRFL08,K09,RSF10,TK10}, $\mathcal{C}$-insulators are formally classified
like superconductors: although they admit an integer classification in two
space dimensions, their weak index $\vec N$ is given by $\mathbb{Z}_2$ valued invariants. However,
despite this formal connection, we find that to predict the presence/absence of zero modes in a
$\mathcal{C}$-insulator requires finer information beyond the weak index in Eq.~\eqref{eq:Ndotb}.
The crucial new ingredient in a $\mathcal{C}$-insulator is the parity of the number of orbitals per
site
\begin{equation}
(-1)^{\eta_{\alpha}}=(-1)^{n_{{\rm orb},\alpha}},
\label{eq:eta}
\end{equation}
where $\alpha$ labels a site.
For $\mathcal{C}$-insulators
with $\eta_{\alpha}=0$ for all the sites $\alpha$ in the unit cell, Eq.~\eqref{eq:Ndotb} works
perfectly just like for their superconducting counterparts. We call these systems {\em even}
$\mathcal{C}$-insulators.
On the other hand, $\mathcal{C}$-insulators with at least one pair $(\alpha,\beta)$ of sites in the
unit cell satisfying $\eta_{\alpha}=\eta_{\beta}=1$ are called {\em odd} $\mathcal{C}$-insulators.
For these odd $\mathcal{C}$-insulators Eq.~\eqref{eq:Ndotb} needs to be modified in an important
way, which is the focus of this work. 

The unique feature of odd $\mathcal{C}$-insulators is that a {\em vacancy} may
cause a change in the zero-mode number by one \cite{PGL06,HZW13}. This is possible if $\mathcal{C}$
acts locally
for every site with $\eta_{\alpha}=1$ (i.e. the operator $\mathcal{C}$ is diagonal for these
orbitals). To understand this let us first consider an ideal odd
$\mathcal{C}$-insulator with a full gap. We assume that $n_{\rm orb}=1$ for a site $i=i_0$ where we
want to create the vacancy. In the first step, we smoothly
switch off all the hopping amplitudes connected to $i_0$ in a manner which preserves
$\mathcal{C}$. Since local $\mathcal{C}$-symmetry forbids a non-zero on-site
potential, a zero-energy state $\psi_0$, which is localized at $i_0$ and self-conjugate
$\mathcal{C}\psi_0 = \pm \psi_0$, must appear once the site $i_0$ is
completely isolated.
However, the only way this zero mode can be generated in
a $\mathcal{C}$-symmetric fashion is by pulling {\it
two} modes to zero, one from the conduction and one from the valence band. Hence, during the
switching-off procedure, two states $\psi_{\pm}$ appear in the gap. Since there is an energy gap
elsewhere, both $\psi_{+}$ and $\psi_-$ are localized in the vicinity of $i_0$. Furthermore,
$\psi_{\pm}$ are conjugate to each other and therefore must be of the form $\psi_{\pm}= \psi_0 \pm
\psi_v$ with $\psi_v$ also self-conjugate $\mathcal{C}\psi_v=\mp\psi_v$. In the second
step, we remove site $i_0$ with its zero-mode $\psi_0$ (also a $\mathcal{C}$-invariant operation
because $\psi_0$ is self-conjugate). We are then left with a single zero mode
$\psi_v$ which is the the vacancy bound state. Adding an even number of orbitals to each site with
$\eta_{\alpha}=1$, this procedure would give an
additional even number of vacancy modes which in general hybridize to finite energies. Hence,
$(-1)^{\eta_{\alpha}}$ predicts the parity of
zero-energy bound states in vacancies of type $\alpha$. This argument is general for all
$\mathcal{C}$-insulators, whether they are topological or not.

We can now see how the existence of these ``non-topological" zero modes prohibits the naive
application 
of Eq.~\eqref{eq:Ndotb}: if we take a dislocation and remove a site from its core,
its zero mode content is changed while its Burgers vector and weak index are unaffected. This is incompatible with
Eq.~\eqref{eq:Ndotb}. Given that there is no meaningful way to determine which of the two dislocation
cores ``contains" the vacancy, can zero modes still be predicted from the weak index? In the
following we will show that with a generalization of Eq.~\eqref{eq:Ndotb}, this is indeed the case.

\begin{figure}[t]
\begin{center}
\includegraphics[width=9cm]{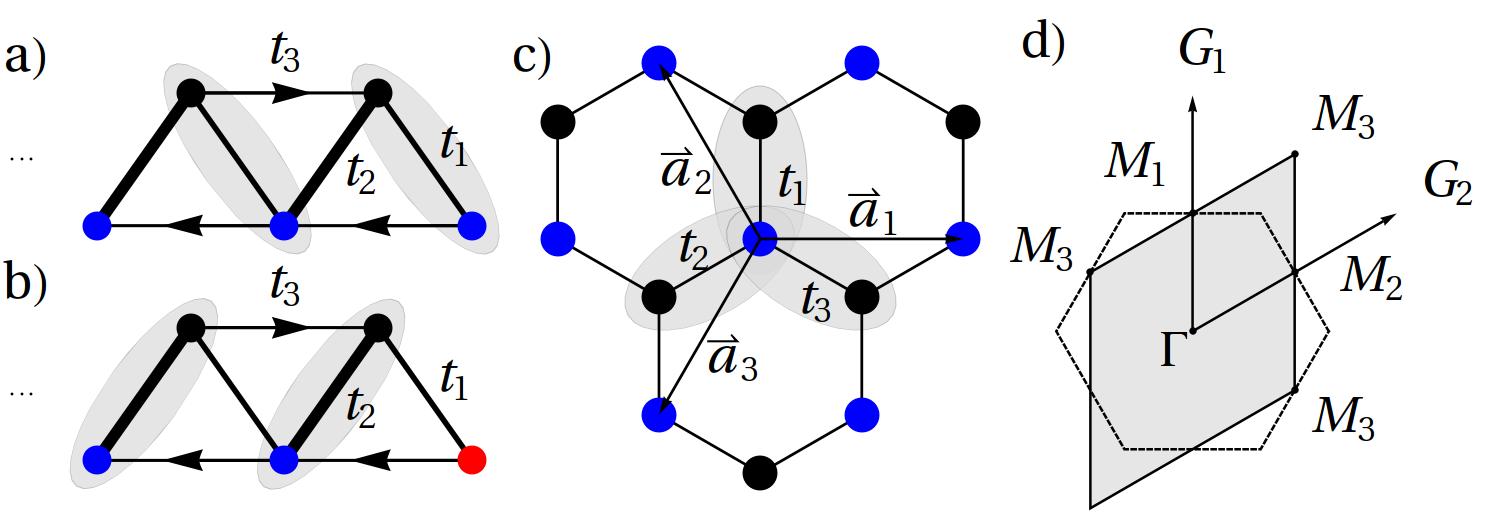}
\caption{a) The 1D model described in the text, with the unit cell enclosing $t_1$ links. Arrows
denote imaginary hoppings. b) The same model with the unit cell enclosing $t_2$ links, which leaves
a broken unit cell (red site) at the edge. The $\mathds{Z}_2$ index computed in (a) is
non-trivial for $t_2>t_1$ in agreement with the presence of an end mode. However, in (b) the index
is zero if $t_2>t_1$. c) Honeycomb lattice in real space. The
three natural choices of unit cells are marked with gray ovals. d) Brillouin Zone and reciprocal
lattice
vectors.}
\label{models}
\end{center}
\end{figure}

\emph{1D model -} 
To warm up let us discuss the zero modes at the ends of an one dimensional odd
$\mathcal{C}$-insulator. In 1D
$\mathcal{C}$-symmetry admits a $\mathbb{Z}_2$ classification. For
concreteness let us consider the following Hamiltonian which is a time-reversal breaking version of
the spinless-polyacetylene model \cite{SSH79,QHZ08,HPB11,DUM11}, depicted in Fig.~\ref{models}(a).
In Fourier space, the bulk Hamiltonian is
\begin{equation}
H(k) = (t_1 + t_2 \cos k)\sigma_x  +  t_2 \sin k \; \sigma_y+2t_3\sin k\;\sigma_z,
\label{eq:H1D}
\end{equation}
where the unit cell of length $a=1$ is chosen such that it encloses the $t_1$ bonds. This
system has a particle hole symmetry
\begin{equation}
U_C^{\dagger} H^{*}(-k) U_C = -H(k), 
\label{eq:PHS}
\end{equation}
with $U_C = \sigma_z$, which is diagonal. The
Hamiltonian Eq.~\eqref{eq:H1D} is in class~$D$ and admits a $\mathds{Z}_2$ topological
invariant which is expressed as the Zak phase \cite{Z89} of the Berry connection of the occupied
band $ A_-(k) = i \left \langle u_-(k)| \partial_k
|u_-(k)\right\rangle $ as \cite{QHZ08}
\begin{equation}
P= \frac{1}{2\pi} \int_{-\pi}^{\pi} \!dk\, A_-(k) \mod 1.
\label{P}
\end{equation}
$P$ is well defined if a gauge-fixing condition for the cell-periodic Bloch functions is employed:
$|u_n(k+2\pi)\rangle=|u_n(k)\rangle$;
indeed, $P$ is invariant under the (remaining) gauge transformations $|u_n(k)\rangle\mapsto
e^{i\beta_n(k)}|u_n(k)\rangle$ with $\beta_n(k)=\beta_n(k+2\pi)\mod{2\pi}$. Furthermore,
particle-hole
symmetry implies $P=-P\mod 1$ and hence $P=0$ or $1/2$ \cite{QHZ08}. As in the theory of
polarization \cite{VKS93}, the Zak phase $P$ indicates if the Wannier center is within the unit cell
($P=0$) or between neighboring unit cells ($P=1/2$) \footnote{Note that a different gauge (called
the
periodic gauge) for the Bloch functions $\psi_{n,k}(x)=e^{ikx}u_{n,k}(x)$ is employed in the theory
of polarization \cite{VKS93}:
$\psi_{n,k+2\pi}(x)=\psi_{n,k}(x)$. This implies a different definition of the Fourier
transformation than the one used to obtain Eq.~\eqref{eq:H1D}. In the periodic gauge, {\em spatial}
symmetries may quantize the value of the Zak phase \cite{HPB11}.}.
Alternatively, $P$ can also be expressed directly in terms of the Hamiltonian at the two
particle-hole invariant momenta (PHIM), $\Gamma=0$ and $M=\pi$. 
For a $2\times2$ Hamiltonian written as $H(k) = \vec f(k) \cdot \vec \sigma$ and with $U_c=\sigma_z$
the
topological invariant is given by
\footnote{This is equivalent to the usual definition with the relative sign of the
Pfaffian at $k=\Gamma,M$, see Suppplementary Materials for more details.} 
\begin{equation}
(-1)^{2 P} = \text{sign}[f_x(\Gamma)]\; \text{sign}[f_x(M)]=\text{sign}(t_1-t_2).
\label{eq:fP}
\end{equation}

The value of $P$ distinguishes between two topological classes of 1D $\mathcal{C}$-insulators.
However, knowing $P$ is not sufficient to deduce the existence of a zero-energy end mode. Let us
look at the right end of a semi-infinite chain [Fig.~\ref{models}(a)] where the last
bond is a $t_1$ link, and suppose that $t_2>t_1$, i.e.~$P=1/2$. Then, we indeed find a zero-energy
end mode which binds a fractional charge 1/2 \cite{QHZ08,HPB11}. But if we remove the last site (so
that the chain terminates with a $t_2$ link) there is no zero mode, because this operation generates
an extra one mod 2. Note neither termination of the chain breaks the particle-hole symmetry (because
$U_C$ is diagonal).

Our inability to predict end states just from $P$ in the 1D chain is thus rooted in the fact that
$\eta_A=\eta_B=1$, cf.~Eq.~\eqref{eq:eta}. To overcome this limitation, we additionally need to know
the number of sites $S_{\alpha}$ left in a broken unit cell at the end of the chain. Then, 
\begin{equation}
n = 2 P + n_b \; \text{mod 2}, \label{1D}
\end{equation}
with $n_b=\sum_{\alpha}S_{\alpha}\eta_{\alpha}$ predicts the $\mathds{Z}_2$ number of zero modes.
Both $P$ and $n_b$ ($S_\alpha$) depend on the unit cell choice but the number $n$ is invariant.
Indeed, if we choose a unit cell enclosing the $t_2$ bonds, a broken unit cell is left at the edge
[Fig.~\ref{models}(b)], but the bulk Hamiltonian is now given by Eq.~\eqref{eq:H1D} with switched
$t_1$ and $t_2$. The eigenstates transform as $|u_n(k)\rangle \mapsto U(k) |u_n(k)\rangle$ with
\begin{equation}
U_{\alpha\beta}(k) = \delta_{\alpha\beta}\left( \delta_{\alpha A} e^{i k}+\delta_{\alpha B} \right).
\label{eq:U}
\end{equation}
This implies a change of $P$ by 1/2 (consistent with the interpretation of $P$ as the
electronic charge center) and Eq.~\eqref{1D} remains invariant. We also note that for the model
Eq.~\eqref{eq:H1D}, the relation Eq.~\eqref{1D} can be
explicitly derived from a Green's function approach \cite{SM,MS11}. 

From a practical point-of-view, Eq.~\eqref{1D} states that a unit cell
choice
which is compatible with the edge, i.e., the unit cell choice where in the open chain there are no
broken unit cells, is convenient to compute the number of zero modes \cite{RH02},
because in that case we simply have $n=2P$.
This fact will prove useful also in higher dimensions.

\begin{figure}[t]
\begin{center}
\includegraphics[width=4.2cm]{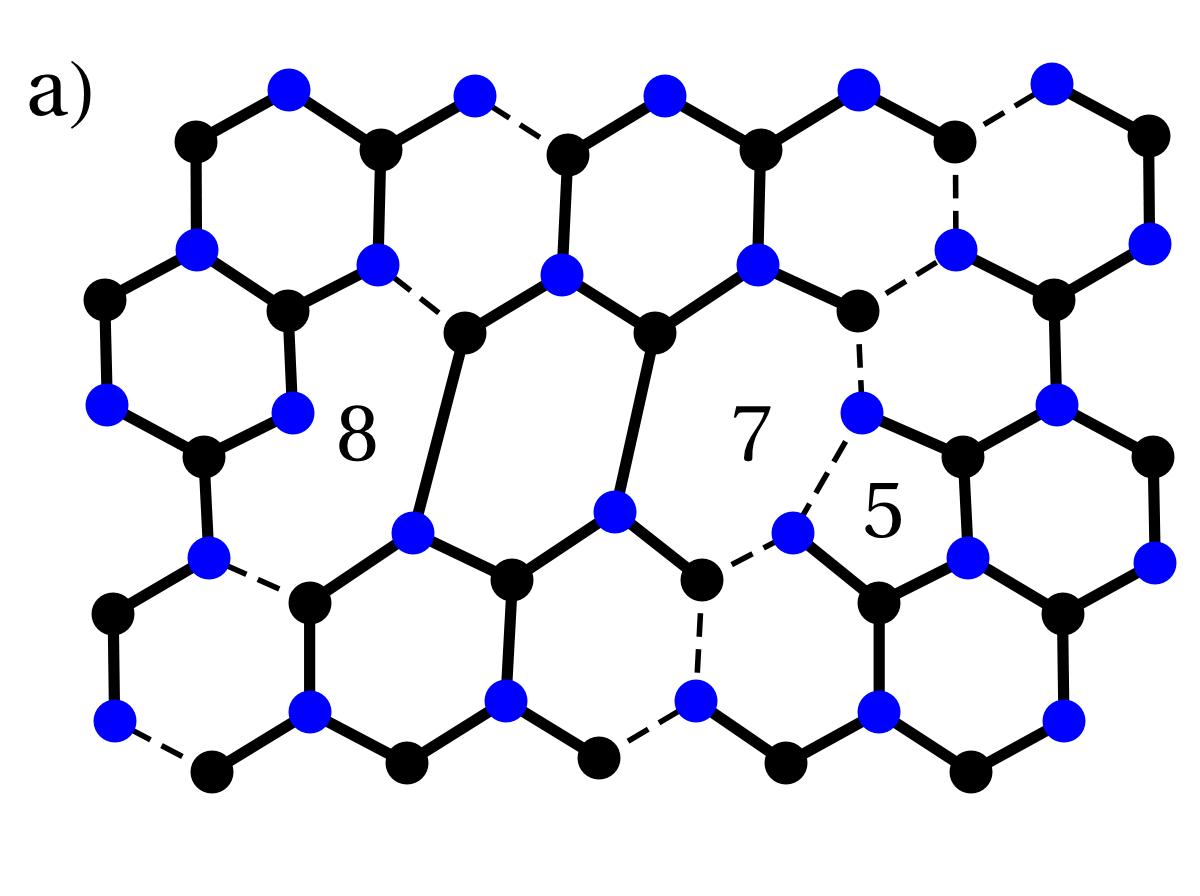}
\includegraphics[width=4.2cm]{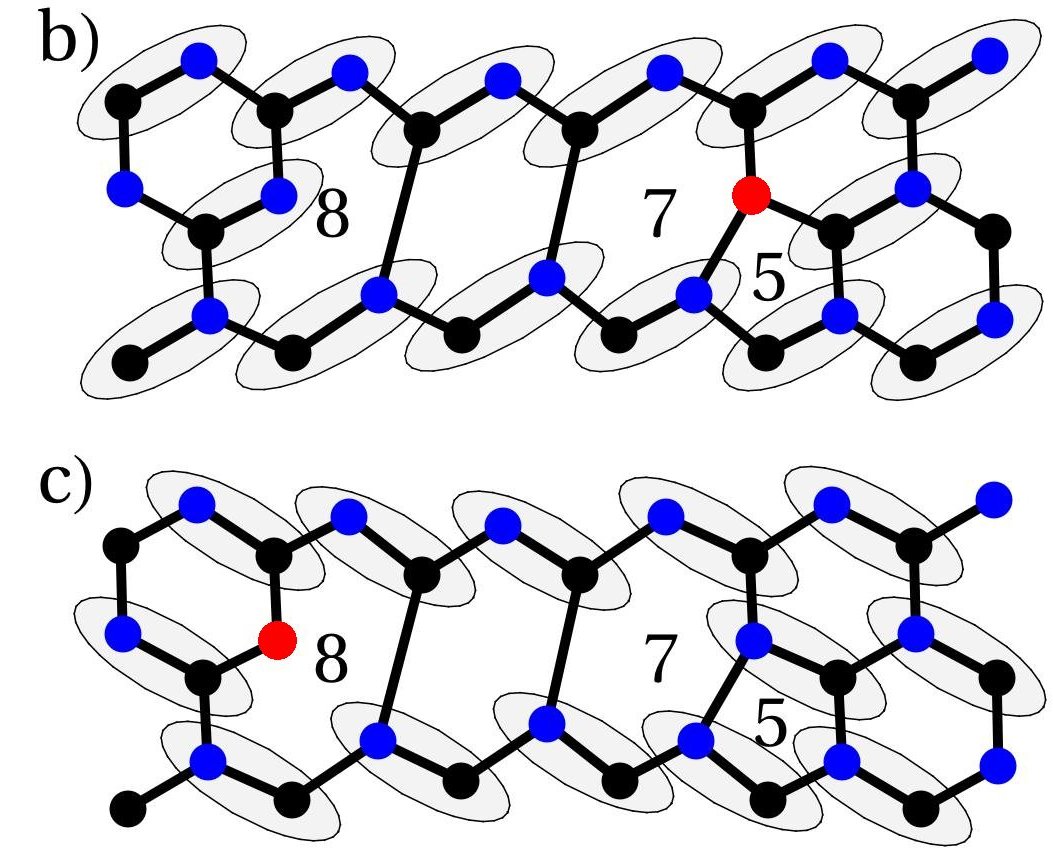}
\caption{a) Honeycomb lattice with a dislocation pair made of 6-8 and 5-7 defects with
burgers vector $\vec b=\pm\vec a_3$.  The 6-8 defect is built by glueing together two zigzag edges
parallel to
$\vec b$ (by the dashed lines) skipping a row in the process. The
5-7 defect is built from two bearded edges in the same way. b) Unit cell 2 is
compatible with the 6-8 defect but leaves a broken unit cell for the 5-7 (red site). c) Unit cell 3
is compatible with the 5-7 defect but leaves a broken unit cell for the 6-8. }\label{dis}
\end{center}
\end{figure}

\emph{Dislocation modes in 2D $\mathcal{C}$-insulators -} We now consider 2D odd 
$\mathcal{C}$-insulators in
the presence of dislocations for which a phenomenon similar to the 1D example can be observed.
Although a class D insulator in 2D has integer classification its weak topological invariants are
$\mathbb{Z}_2$ valued. This is because the weak invariants are determined by the topological index
in one dimension less when one component of the momentum is fixed at a particle-hole symmetric
value.
In the following, we illustrate the generic problem by considering dislocations in the honeycomb
lattice,
well studied in graphene \cite{EHB02,CBJ08,LJV09,MPF10}. In this system, two dislocations with
identical Burgers vector can differ in their core structure. The two possible cores
\cite{EHB02},
shown in Fig.~\ref{dis}(a), are the pentagon-heptagon (5-7) and the hexagon-octagon (6-8) core. 
The importance of the core structure has been realized earlier for gapless graphene
\cite{CBJ08,MPF10}, where the 6-8 defect binds a zero mode (to which no topological meaning has been
assigned so far) but the 5-7 does not. 

Here, we discuss an odd $\mathcal{C}$-insulator (topological) model on the honeycomb lattice [see
Fig.~\ref{models}(c)] which admits a
topological characterization and which shows a similar dependence on the microscopic structure of
the dislocation core.
In the Supplementary materials \cite{SM}, we discuss further examples of even and odd 
$\mathcal{C}$-insulators \cite{W08,CH94,BHZ06,FIH13,KM05} which support our main conclusion.
Choosing the unit cell which encloses
the bond $t_1$, we define the bulk Hamiltonian as
\begin{align}
&H(\vec k) = \sigma_x \left(t_1 + t_2 \cos k_1 + t_3 \cos k_2 \right) +  \label{ham}
\\ 
\sigma_y &\left(t_2 \sin k_1 +t_3
\sin k_2 \right) + t' \sigma_z ( \sin k_1+\sin k_2+\sin k_3), \nonumber
\end{align}
where $t_1,t_2,t_3$ are the three nearest neighbor hoppings, $k_i=\vec k \cdot \vec a_i$, and the
last term is known as the
Haldane mass \cite{H88}. This model has particle-hole symmetry with $U_C = \sigma_z$. Depending on
the value of $t'$, it can either be a Chern insulator or a trivial insulator in the D class. For
example, if $t_1=t_2=t_3$ and $t'\neq 0$, a Chern insulator with Chern
number $C=\pm1$ is realized \cite{H88}.
If $t'\rightarrow0$ and one of the first-neighbor amplitudes exceeds twice the magnitude of the
others
$|t_{l}|>2|t_{m\neq l}|$ (breaking $C_6$ rotational symmetry to $C_2$), the Dirac points move from
$K$ and $K' $ to the $M_l$ point where they annihilate and open a gap which results in a trivial
insulator. However, as discussed earlier, even a trivial 2D $\mathcal{C}$-insulator can have zero
modes in the dislocation core. In later discussions we will refer to this model as
``dimerized model in bond $t_l$". 

Analogously to the 1D chain discussed above, $\eta_A=\eta_B=1$ implies a dependence of $\vec N$ on
the unit cell choice. Not surprisingly we find Eq.~\eqref{eq:Ndotb} requires a modification: 
the dislocation carries a non-trivial zero mode if
\begin{equation}
\frac{\vec{N}\cdot\vec{b}}{\pi}+n_b=1 \mod 2.
\end{equation}
If we compute the weak index
with a unit cell choice that is compatible with the defect,
Eq.~\eqref{eq:Ndotb} works perfectly.
 
\begin{figure}[t]
\begin{center}
\includegraphics[width=8.7cm]{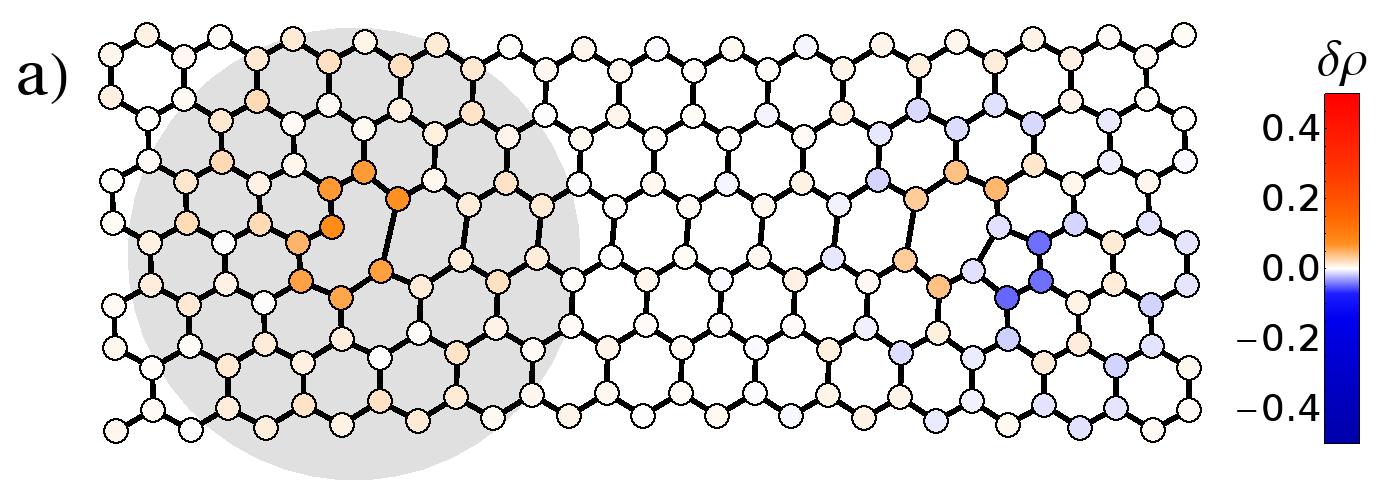}
\includegraphics[width=8.7cm]{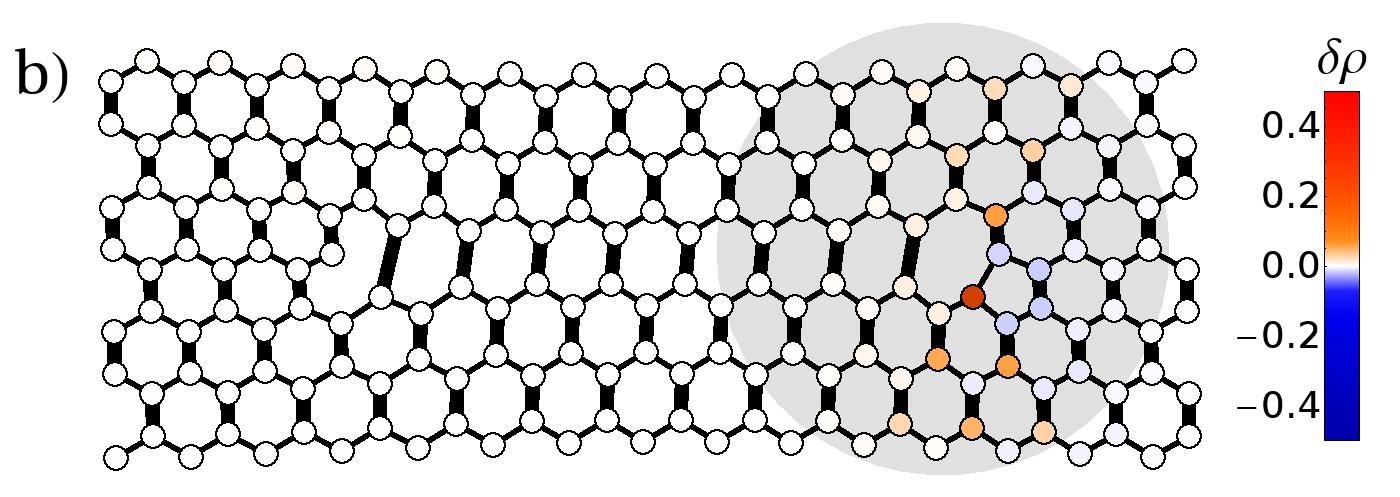}
\caption{Charge distribution for occupied valence band and zero mode in the presence of a
dislocation pair (6-8 and
5-7) with $\vec b =\pm \vec a_1$. Only a portion of the lattice used for the computation is shown.
a) Haldane model. The 6-8 has a zero mode and binds 1/2 charge within the gray disk. The charge
around the 5-7 adds up to to zero. b) $t_1$-dimerized model. Now the 5-7 has a zero-mode and
binds 1/2 charge, while the 6-8 is trivial. See \cite{SM} for more details.}
\label{dipoles}
\end{center}
\end{figure}

By writing $H(\vec k) = \vec \sigma \cdot \vec f(\vec k)$, the weak index is evaluated from
$f_x(k_i)$ with
$k_i$ the PHIM. These are $\Gamma = (0,0)$, $M_1 =
2\pi /\sqrt{3}a (0,1)$, $M_2 = 2\pi /\sqrt{3}a (-\sqrt{3}/2,-1/2)$ and $M_3 = 2\pi /\sqrt{3}a
(\sqrt{3}/2,-1/2)$, and are illustrated in Fig.~\ref{models}(d). Along the lines
$M_1\rightarrow M_3 \rightarrow M_1$ and $M_2\rightarrow M_3 \rightarrow M_2$, $H(\vec k)$ reduces
to 1D $\mathcal{C}$-insulators for which the $\mathds{Z}_2$ index can be computed similar to
Eq.~\eqref{eq:fP}, 
\begin{align}
(-1)^{\nu_1} = \text{sign}[f_x(M_1)]\; \text{sign}[f_x(M_3)], \\
(-1)^{\nu_2} = \text{sign}[f_x(M_2)]\; \text{sign}[f_x(M_3)].
\label{wi1}
\end{align}
The weak index is then defined as
$
\vec N = 1/2(\nu_1 \vec G_1 +\nu_2 \vec G_2).
$
There are three natural choices of unit cells, each of which encloses one of the nearest-neighbor
links, see
Fig.~\ref{models}(c). For
unit cell 1, for example, we have $f_x(\Gamma) =t_1 + t_2 + t_3$, $f_x(M_1) =t_1 - t_2 - t_3$,
$f_x(M_2) =t_1 - t_2 + t_3$ and $f_x(M_3) =t_1 + t_2 - t_3$.
These expressions change with the choice of unit cell. For the Haldane model with $i$-type of unit
cell choice, the
weak index turns out to be
\begin{equation}
\vec N = \frac{1}{2} \vec G_i, \label{hal}
\end{equation}
while for the $t_i$-dimerized model with $j$-type unit cell the weak index is 
\begin{equation}
\vec N = \frac{1}{2} \epsilon_{ijk} \vec G_k. \label{nem}
\end{equation}
Using Eqs.~\eqref{hal} and \eqref{nem} and the possible unit cell tilings for 6-8 and 5-7
defects in Fig.~\ref{dis}, one determines that the Haldane model always has a zero mode in the 6-8
defect but not in the 5-7 (this explains recent statements about dislocation
modes in this model \cite{JMS12,SMJ13}). If we denote the nearest-neighbour vector for
the dimerized bond $t_i$ as $\vec \delta_i$, the dimerized model with $\vec b \cdot \vec \delta_i =
0$
has a zero mode in the 6-8, but if $\vec b \cdot  \vec \delta_i \neq 0$ then it has a zero mode in
the 5-7. This demonstrates that zero modes in the 6-8 dislocation are neither accidental nor
associated with a
``dangling bond``. They are topological and determined by the weak index and $n_b$ - as are the
zero-modes of the 5-7
dislocations. We have explicitly checked these statements by implementing a dislocation dipole in a
tight-binding lattice with periodic boundary conditions. Bound fractional charges were
determined by examining the sum of $\delta \rho_i = (\sum_{occ} |\psi_i|^2) -1/2$ over a
sufficiently large disk enclosing the defects. The results are shown in Fig.~\ref{dipoles}.

The existence of dislocation zero modes is intuitively understood by
considering the cut-and-glue construction of a dislocation with Burgers vector $\vec b$
\cite{RZV09}. On the honeycomb lattice, the 6-8 and 5-7 dislocations are constructed by glueing
together two zigzag or two bearded edges, respectively, while skipping a row in the process, see
Fig.~\ref{dis}. A dislocation mode can then be viewed as a non-trivial bound state carried by the
mass kink which appears if two edges having a single mode crossing $E=0$ at momentum $k = \pi$ are
coupled \cite{SSH79,LZX07,R10}. This precisely happens when $\vec N \cdot \vec b = \pi$ if $\vec N$
is computed with a unit cell compatible with the edge. Using Eqs.~\eqref{hal} and \eqref{nem},
the locations of the zero-energy crossings are easily determined. This gives results consistent with
previous studies of zigzag and bearded edges \cite{SM,DUM11,TM13}.

{\em Dislocation modes in the gapless case -} So far we have focused on insulators, but our results
generalize to $\mathcal{C}$-invariant nodal semimetals 
\cite{RH02,DUM11,SF10,MGV07,HKV11,SR11,WL12}. So long as the node does not fall on the relevant one
dimensional lines in $k$-space which are used to compute the weak index, all the earlier analysis
follows \cite{SF10}. In our case, the model \eqref{ham} with
$t_1=t_2=t_3$ and $t'\rightarrow0$ describes (spinless) graphene. Because the spectrum remains fully
gapped along the lines $M_1\rightarrow M_3 \rightarrow M_1$ and $M_2\rightarrow M_3 \rightarrow
M_2$, the weak index preserves its meaning. The edge state count
at $k=\pi$ and the dislocation modes thus remain unaltered, but now coexist with bulk zero energy
states located at the nodes. We conclude that the zero mode at a 6-8 dislocation in graphene
\cite{CBJ08,MPF10}, for
which experimental evidence exists \cite{CKF09}, has a topological origin and is protected by
particle-hole symmetry. Finally, a weak index can also be defined for nodal 3D systems (Weyl
semimetals) which allows to deduce, e.g.,~1D modes in screw dislocations \cite{IT11}.

\emph{Discussion -} 
The finer structure of odd $\mathcal{C}$-insulators obtained in this work relies on the fact that
vacancy zero modes appear if $\eta_{\alpha}=1$. This implies a unit-cell
dependence of the weak index and a core dependence of dislocation modes. It is also possible to turn
the argument around and to deduce the vacancy mode from the core dependence of the dislocation
modes. For this purpose, consider a pair of dislocations with opposite Burgers vectors $\pm
\vec b$ and imagine to mechanically deform the lattice to bring the two defects on top of each
other. Because the total Burgers vector vanishes, no translational
holonomy is produced, but the two defects can annihilate into a perfect lattice or into a vacancy
depending on their type
\cite{CBJ08}. The second case precisely occurs when no unit cell choice can be found that tiles the
dislocation dipole as a whole (such as the dipoles in Fig.~\ref{dipoles}). Then, if only one type of
dislocation binds a zero mode but not the other, also an isolated vacancy must bind a zero mode. We
emphasize, however, that there is no particular dislocation, 6-8 or 5-7, that acts as a vacancy by
itself, as we have seen from the difference between the Haldane and the dimerized model.

Let us also discuss how $\eta_{\alpha}=1$ can affect the presence of disclination modes.
Disclinations are defects which
are characterized by a {\em rotational} holonomy and it was shown that such defects can give rise to
non-trivial bound states \cite{LC00,LC04,RL13,RCM13,TH13}. On the honeycomb lattice, 120$^{\circ}$
disclinations are compatible with the $\mathcal{C}$-symmetry and there are three different types:
the square, the 5-5 and the 6-6 disclinations \cite{SM}. The Haldane
model has a zero mode in the square disclination \cite{RL13} and the 5-5 disclination but no zero
mode in the 6-6 disclination. The latter two disclination types are related to each other by the
removal of a
single site, (creation of a vacancy) and because $\eta_A=\eta_B=1$ they differ by the presence of a
zero mode.

Finally, we note that the dimerized model is readily available in artificial graphene experiments
\cite{GMK12}. We also expect that 
dislocations can be built in these systems to test our predictions. Furthermore, we anticipate that
our results are also applicable
to 3D bipartite lattices \cite{FKM07,TM13}, where we expect a core-dependence of 1D dislocation
modes.
\acknowledgements
The authors thank T. Chen for useful discussions. F.\ de J.\ acknowledges financial support  from
the ``Programa Nacional de Movilidad de Recursos Humanos" (Spanish MECD) and A.R. from the Swiss
National Science Foundation.

\bibliography{defects}


\pagebreak
\onecolumngrid
\vspace{0.2in}
\begin{center}
{\bf \large Bulk-Defect Correspondence in Particle-Hole Symmetric Insulators and Semimetals -
Supplementary materials}
\end{center}
\vspace{0.1in}

\renewcommand{\thetable}{S\Roman{table}}
\renewcommand{\thefigure}{S\arabic{figure}}
\renewcommand{\thesubsection}{S\arabic{subsection}}
\renewcommand{\theequation}{S\arabic{equation}}

\setcounter{secnumdepth}{1}
\setcounter{equation}{0}
\setcounter{figure}{0}
\setcounter{section}{0}

\section{Topological invariant}
The general procedure \cite{K01} to compute the 1D topological invariant in class $D$ requires to
express $H$
in a basis in which $U_c=\mathcal{I}$ (the Majorana basis), where $U_c$ is the unitary part of the
charge conjugation operator $\mathcal{C}=U_c\mathcal{K}$ ($\mathcal{K}$ denotes the complex
conjugation)
\begin{equation}
U_c H^*(-k)U^{\dagger}_c = -H(k). \label{ph}
\end{equation}
In
general, this is achieved by a unitary
transformation $H_M = U H U^{\dagger}$. Due to the antiunitarity of $\mathcal{C}$, the unitary part
of $\mathcal{C}$ transforms as $U_c
\rightarrow U U_c U^{T}$. In this basis, the Hamiltonian at the PHIM is purely imaginary and
antisymmetric, $H_M=iA$, $A^{T} =- A$, and the invariant is expressed in terms of the Pfaffian at
these points 
\begin{equation}
(-1)^{2P} = \text{sign}[\text{Pf}(A(\Gamma))]\text{sign}[\text{Pf}(A(M))].
\end{equation}
This is particularly simple for $2\times2$ Hamiltonians for which the Pfaffian is simply the number 
$A_{12}=-A_{21}$.  

We can now apply this method to any Hamiltonian of the form $H = f_x \sigma_x +f_y \sigma_y + f_z
\sigma_z$ with $U_c=\sigma_z$. Note particle-hole symmetry implies $f_x(-k)=f_x(k)$ and
$f_{y,z}(-k) = - f_{y,z}(k)$, so that at PHIM only $f_x$ is nonzero. We now go to the Majorana
basis with $U=\text{diag}(1,i)$, so that $U_c = U \sigma_z U^{T} = \mathcal{I}$ and
\begin{equation}
H_M(k) = U H U^{\dagger} =  f_x \sigma_y - f_y \sigma_x + f_z \sigma_z.
\end{equation}
At the PHIM, $H$ is proportional to $\sigma_y$ (imaginary and antisymmetric) and the Pfaffian is
simply $-f_x$, so that the topological invariant reduces to 
\begin{equation}
(-1)^{2P} = \text{sign}[f_x(\Gamma)]\; \text{sign}[f_x(M)],
\label{eq:fPSM}
\end{equation}
as quoted in the main text. 

In two dimensional models, the 1D lines at the edges of the Brillouin zone can be considered as 1D
models. The corresponding 1D invariants can be computed and determine the weak index. In the
honeycomb
lattice, the PHIM are $\Gamma$ and the three $M$ points, and following standard notation we have 
\begin{align}
(-1)^{\nu_1} = \text{sign}[\text{Pf}(A(M_1))]\; \text{sign}[\text{Pf}(A(M_3))], \\
(-1)^{\nu_2} = \text{sign}[\text{Pf}(A(M_2))]\; \text{sign}[\text{Pf}(A(M_3))],
\label{weakhon}
\end{align}
while for the square lattice the PHIM are now $\Gamma$, $X$, $Y$ and $M$, and we have 
\begin{align}
(-1)^{\nu_1} = \text{sign}[\text{Pf}(A(X))]\; \text{sign}[\text{Pf}(A(M))], \\
(-1)^{\nu_2} = \text{sign}[\text{Pf}(A(Y))]\; \text{sign}[\text{Pf}(A(M))]. \label{wisq}
\end{align}
The weak index is then computed as $\vec N = 1/2 (\nu_1G_1 + \nu_2 G_2)$. For $2\times2$ models with
$U_c=\sigma_z$ we can directly replace the Pfaffian by $f_x$ as in one dimension.  

\section{End states in 1D model: Green function approach}

In this section, we explicitly derive the relation Eq.~\eqref{1D} for the 1D model
Eq.~\eqref{eq:H1D} using the Green's function approach \cite{MS11}. The Berry connection for model
Eq.~\eqref{1D} is given by
\begin{equation}
A_k=\langle u_-(k)|(i\partial_k)|u_-(k)\rangle=\frac{n_x\partial_kn_y-n_y\partial_kn_x}{2(1+n_z)}, \
\end{equation}
where the cell-periodic Bloch function $u_-(k)$ of the lower band of the $\mathcal{C}-$symmetric
two-band Hamiltonian $H_0(k)=\vec f(k)\cdot\vec\sigma$ is
\begin{equation}
u_-=\frac{1}{\sqrt{2}}
\begin{pmatrix}
-\frac{\sqrt{n_x^2+n_y^2}\sqrt{1-n_z}}{n_x+in_y}\\
\sqrt{1+n_z}
\end{pmatrix},\quad \vec n(k)=\frac{\vec f(k)}{|\vec f(k)|}.
\end{equation}
Because $n_z(-k)=-n_z(k)$, one shows that the Zak phase Eq.~\eqref{P} is independent of $n_z(k)$ and given by:
\begin{equation}
2P=\cos^2\left(\frac{1}{2}\int_0^{2\pi}A_k\;dk\right)=\cos^2\left(\frac{1}{2}\int_0^{2\pi}A^{xy}
_k\;dk\right)=\Theta(t_2-t_1),
\end{equation}
where $A^{xy}(k)=(n_x\partial_kn_y-n_y\partial_kn_x)/[2(n_x^2+n_y^2)]$ is the Berry connection for $n_z\equiv0$.

The general procedure for using Green's function to proof the existence of end states is as follows
\cite{MS11}. First, consider the system with periodic boundary conditions, described by the
Hamiltonian $H_0$. Now add an ``impurity potential" $V$, which cuts all the links crossing a given
fictitious boundary and hence transforms the periodic system into an open system $H=H_0+V$. The
single-particle Green function $G(E)=(E-H+i\delta)^{-1}$ of the open system is then related to the
single-particle Green function $G^{0}(E)=(E-H_0+i\delta)^{-1}$ of the periodic system by
\begin{equation}
G(E)=G^{0}(E)[{\boldsymbol 1}-VG(E)]^{-1}.
\end{equation}
The presence of end modes are signaled by additional zero-energy poles in $G(E)$, i.e.~$\det G^{-1}(E=0)=0$.
We distinguish two cases: first, a cut between unit cells and second, a cut within a unit cell.
\subsection{Cut between unit cells}
Let's assume to cut the system between unit cell 0 and unit cell 1. This corresponds to an impurity potential
\begin{equation}
V=\begin{pmatrix}
0&V_{01}\\
V_{01}^{\dag}&0
\end{pmatrix},\quad
V_{01}=\begin{pmatrix}
-it_3&0\\
-t_2&it_3
\end{pmatrix},
\end{equation}
in the basis $(A0,B0,A1,B1)$. We define the (independent) components of the local Green function of the periodic system at $E=0$ as follows
\begin{equation}
G^0(E=0)=\begin{pmatrix}
G^0_{00}&G_{01}^{0}\\
G_{10}^0&G_{11}^0
\end{pmatrix},\quad
G^0_{00}=G_{11}^0=\begin{pmatrix}
0&g_{AB}\\
g_{AB}&0
\end{pmatrix},\quad
G_{01}^0=(G_{10}^0)^{\dag}=
\begin{pmatrix}
ig_{A0A1}&g_{A0B1}\\
g_{B0A1}&-ig_{A0A1}
\end{pmatrix}.
\end{equation}
The condition for zero-energy end states, $\det G^{-1}(E=0)$, reduces to
\begin{equation}
[1+ t_2g_{B0A1}+2t_3 g_{A0A1}+t_3^2\left(g_{A0A1}^2+g_{AB}^2-g_{A0B1}g_{B0A1}\right)]^2=0.
\end{equation}
Thus, if the above condition is satisfied, two zero-modes appear, one on each end of the chain.
The components of Green's function can be evaluated using complex analysis and one finds
\begin{equation}
1+ t_2g_{B0A1}+2t_3 g_{A0A1}+t_3^2\left(g_{A0A1}^2+g_{AB}^2-g_{A0B1}g_{B0A1}\right)=\left(1+\frac{t_1}{\sqrt{t_1^2+4t_3^2}}\right)(1/2-P).
\end{equation}
Indeed, there is a zero-energy end state if $P=1/2$, i.e.~$t_2>t_1$.
\subsection{Cut within unit cell}
The procedure is analogous if we cut within a unit cell, say unit cell 1. The impurity potential is then given by 
\begin{equation}
V=\begin{pmatrix}
0&V_{11}\\
V_{11}^{\dag}&0
\end{pmatrix},\quad V_{11}=\begin{pmatrix}
it_3&0\\
-t_1&-it_3
\end{pmatrix},
\end{equation}
in the basis $(B0A1B1A2)$. The independent components of the local Green function of the periodic system at $E=0$ are
\begin{equation}
G^0(E=0)=
\begin{pmatrix}
0&g_{B0A1}& -ig_{A0A1}& g_{B0A2}\\
g_{B0A1}&0&g_{AB}&ig_{A0A1}\\
ig_{A0A1}&g_{AB}&0&g_{B0A1}\\
g_{B0A2}&-ig_{A0A1}&g_{B0A1}&0
\end{pmatrix}.
\end{equation}
The condition for a pole at $E=0$ is
\begin{equation}
1+ t_1g_{AB}+2t_3 g_{A0A1}+t_3^2\left(g_{A0A1}^2+g_{B0A1}^2-g_{AB}g_{B0A2}\right)=\left(1+\frac{t_2}{\sqrt{t_2^2+4t_3^2}}\right)P.
\end{equation}
Thus, there is a zero-energy end state if $P=0$ or $t_1>t_2$. 

The results of cutting between or within the unit cell can be summarized to yield the condition
\begin{equation}
1=2P+n_b\mod 2,
\end{equation}
for a zero-energy end mode. Here, $n_b$ is the number of broken unit cells arising from the cut. This is Eq.~\eqref{1D} of the main text.

\section{Honeycomb lattice models}
\subsection{Single-orbital honeycomb lattice model}
The single-orbital model on the honeycomb lattice is given in Eq.~\eqref{ham} of the main text.  
Here, we present some more details on the weak index computation, the edge spectrum and the
calculations of the dislocation bound charge.

The weak indices of this $2\times2$ model can be computed from $f_x$ in Eq.~\eqref{ham} at the
different PHIM. The values of $f_x$ for the different unit cells are shown in Table~\ref{table}(a).
Substituting $t_1=t_2=t_3=t$ for the Haldane model, or $t_{i=j} = (2+\epsilon)t$,
$t_{i\neq j} = t$ for dimerization in $t_j$ in Eqs.~\eqref{wi1}, we find the weak indices in
Table~\ref{table}(b), which can be compactly encoded in expressions~\eqref{hal} and~\eqref{nem} in
the
main text.

Next we discuss the relation between edge states and weak index for the present models. In general,
edges in a ribbon geometry will have a single mode (per edge) at $k=\pi$ in the one-dimensional BZ
if $\vec N \cdot \vec a = \pi$ with $\vec a$ a translation vector of the ribbon (parallel to the
edge) and with $\vec N$ the weak index computed with a unit cell that tiles the edge. This is
equivalent to the condition for zero modes in dislocations, as can be seen from the edge glueing
construction. 

Consider first the Haldane model, which has a single edge state (due to $C=\pm1$), and has weak
index given by Eq.~\eqref{hal}. Consider an edge paralell to $\vec a_1$. A bearded edge is tiled by
unit cell 1, for which $\vec N = 1/2 \vec G_1$, so $\vec N \cdot \vec a_1 = 0$ and there are no edge
states at $k=\pi$. Zigzag edges are tiled by unit cells 2 and 3, for which $\vec N \cdot \vec a_1 =
\pm \pi$ and an thus edge at $k=\pi$ should be present. This is shown by an explicit computation in
Fig.~\ref{pband}(a).

In the $t_i$-dimerized model $C=0$ and Eq.~\eqref{nem} now determines the weak index. With a 
bearded edge paralell to $\vec a_1$ (unit cell 1), we have $\vec N \cdot \vec a_1 =\pm \pi$ for
dimerization in $t_2$ and $t_3$, which give a zero mode at $k=\pi$ (and another one at $k=0$ because
$C=0$), but $\vec N \cdot \vec a_1 =0$ for dimerization in $t_3$ and no zero mode is present. A
zigzag
edge has the opposite behaviour. Note that with time reversal symmetry, the model is actually in
class BDI and there are in fact flat bands at the edges but a small Haldane mass will make them
dispersive without changing the weak index. This model has also been described in Ref.~\cite{TM13}
where plots of the spectrum can be found and are consistent with this discussion. 

\begin{figure}[t]
\begin{center}
\includegraphics[width=8.7cm]{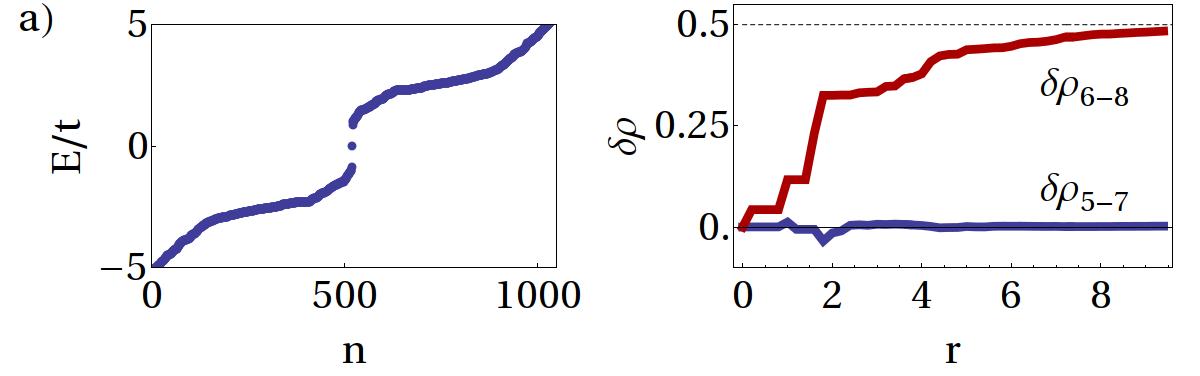}
\includegraphics[width=8.7cm]{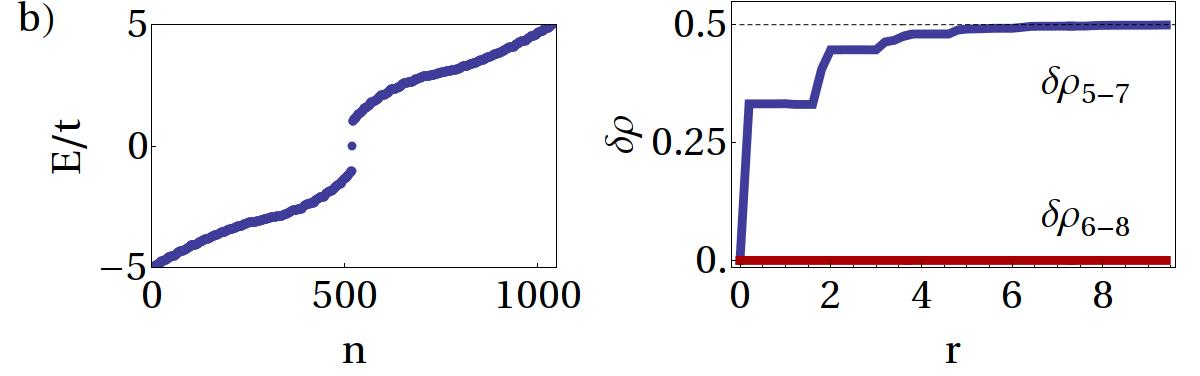}
\caption{a) Left: Spectrum of the Haldane model for $t'=t$, for a larger version of the lattice in
fig. \ref{dipoles}(a) where defects are separated by 20$a$. Right: Integrated charge in a circle of
radius $r$ centered at the 5-7 (blue curve) and 6-8 (red curve) defects. b) The same for the $t_1$
dimerized model in Fig.~\ref{dipoles}(b).}\label{schoneycomb}
\end{center}
\end{figure}

Finally, these honeycomb lattice models were implemented in a tight-binding lattice with periodic
boundary conditions and a pair of dislocations of opposite Burgers vectors, as shown in Fig.
\ref{dipoles} in the main text. We now provide numerical evidence of the presence of zero modes and
the quantization of the fractional charge to 1/2. To provide better precision, we implemented the
model in a larger lattice where the distance between dislocations is 20 lattice constants.
Fig. \ref{schoneycomb} shows the spectra for the Haldane and dimerized model, where the presence of
a single zero mode is evident. It should be noted that the dimerized model has a zero mode in the
5-7 defect, which has a hopping term connecting two sites in the same sublattice and locally breaks
particle-hole symmetry. The energy of the zero mode however remains exactly zero within numerical
precision, but this exactness is not expected to survive disorder.

To probe the fractional charge bound to the defects, we computed $\delta \rho_i = \sum_{E<=0}
|\psi_{E,i}|^2$ and integrated it in a disk of radius $r$ around the defects. This is plotted
also in Fig. \ref{schoneycomb}, where it is seen that the value of the fractional charge converges
to 0 or 1/2 within a few lattice constants from the defects. 

\begin{center}
\begin{table}
\begin{tabular}{|c|c|c|c|}\hline
\multicolumn{4}{|c|}{$f_x$}\\ \hline
\; & Unit cell 1 & Unit cell 2 & Unit cell 3 \\ \hline
$f_x(\Gamma)=$ & $t_1+t_2+t_3$ & $t_1+t_2+t_3$ & $t_1+t_2+t_3$\\ \hline
$f_x(M_1)=$ & $t_1-t_2-t_3$ & $-t_1+t_2+t_3$ & $-t_1+t_2+t_3$\\ \hline
$f_x(M_2)=$ & $t_1-t_2+t_3$ & $-t_1+t_2-t_3$ & $t_1-t_2+t_3$ \\ \hline
$f_x(M_3)=$ & $t_1+t_2-t_3$ & $t_1+t_2-t_3$ & $-t_1-t_2+t_3$ \\\hline
\end{tabular}
\begin{tabular}{|c|c|c|c|}\hline
\multicolumn{4}{|c|}{($\nu_1$,$\nu_2$)}\\ \hline
Model & Unit cell 1 & Unit cell 2 & Unit cell 3 \\ \hline
Haldane & (1,0) & (0,1) & (1,1) \\ \hline
 $t_1$-dimerized & (0,0) & (1,1) & (0,1)\\ \hline
 $t_2$-dimerized & (1,1) & (0,0) & (1,0)\\ \hline
 $t_3$-dimerized & (0,1) & (1,0) & (0,0) \\\hline
\end{tabular}
\caption{Left: Values of $f_x(k)$ at the PHIM for different unit cells in the honeycomb lattice
model. Right: Weak indices as computed from Eq.~\eqref{weakhon} for different models.}\label{table}
\end{table}
\end{center}
\subsection{Two-orbital honeycomb lattice model}
We next consider a two-orbital model on the honeycomb lattice which uses the $p_x$ and $p_y$
orbitals \cite{W08}. Similar to the Haldane model, the planer $p$-orbital model also realizes
non-trivial Chern insulators.

\begin{figure}[t]
\begin{center}
\includegraphics[width=7cm]{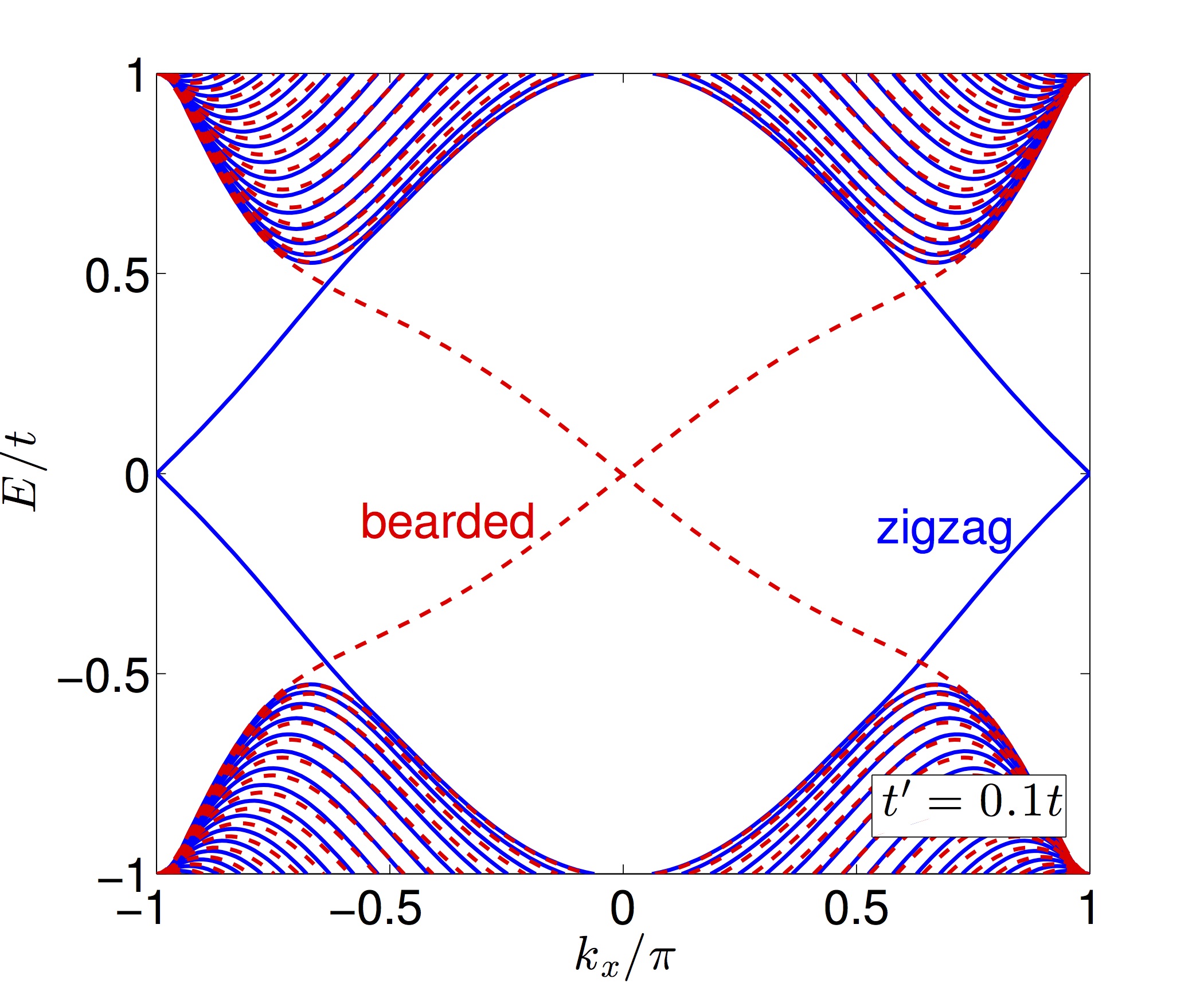}
\includegraphics[width=7cm]{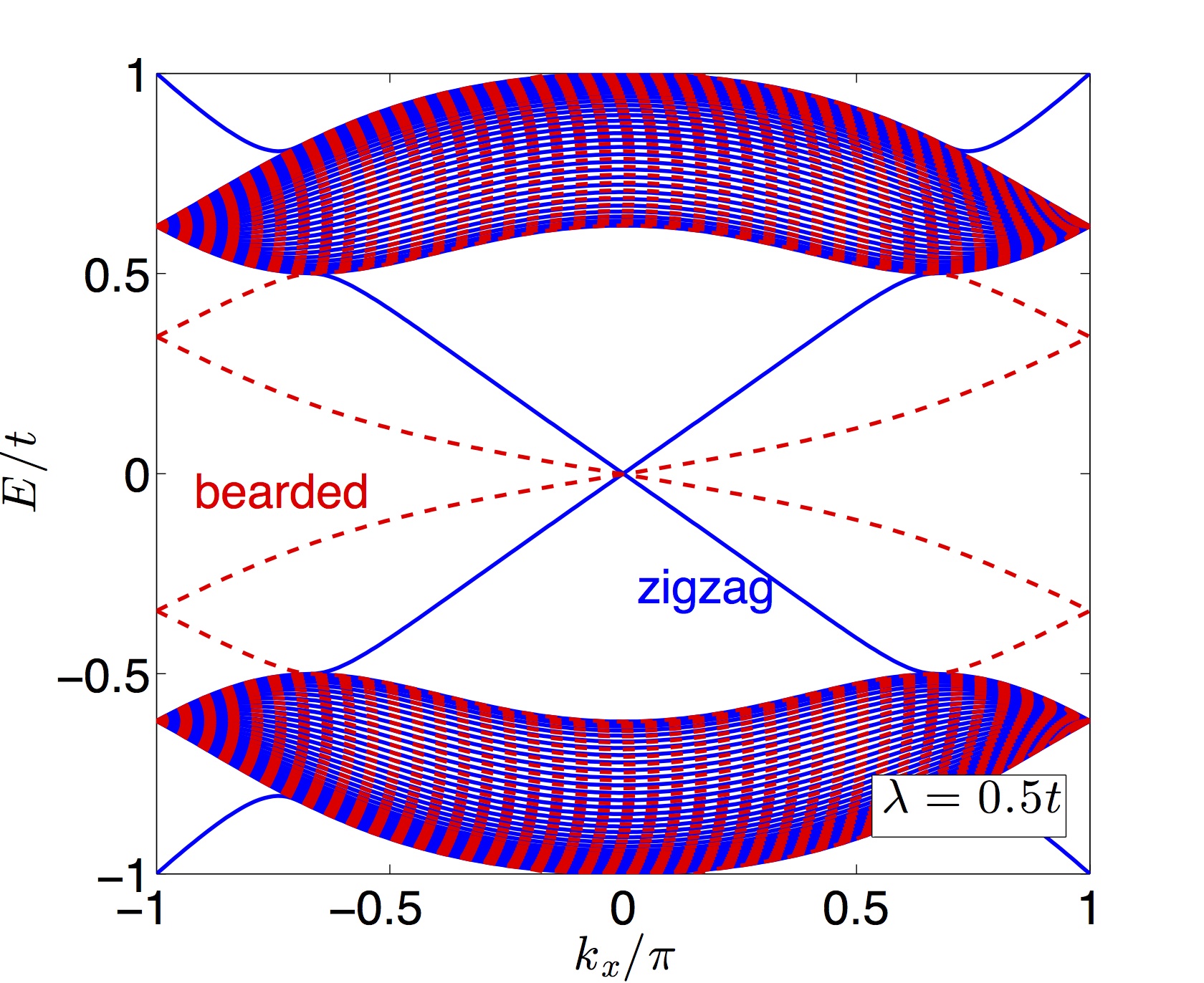}
\caption{Left: Spectrum of the Haldane model in a ribbon geometry
with zigzag and
bearded edges for $t'=0.1t$. Edge modes cross at $k=\pi$ for zigzag but at $k=0$ for bearded, in
agreement with
weak indices computed with unit cells that tile the edges. Right: Spectrum of the p-band model for
$\lambda=0.5t$ and both types of edges. Now in both cases the edge states cross at $k=0$,
in agreement with a trivial weak index.}\label{pband}
\end{center}
\end{figure}
Since this model has two orbitals per site $\eta_A=\eta_B=0$. This implies, first of all, that
there are no zero modes in vacancies in this system.
Moreover, dislocations of either type (5-7 or 6-8) must now carry the same zero mode content, since
the removal of a site (to go from one to the other) does not change the zero mode content. And
finally this also implies that the weak index must now be independent of the unit cell choice. We
now prove this last statement explicitly. For a unit cell centered around $t_1$ bonds as the one
used in the main text, the Hamiltonian in the basis $\psi = (p_x^A, p_y^A, p_x^B, p_y^B)^T$ takes
the
form 
\begin{equation}
H = \left( \begin{array}{cc}
\Omega & t_1 + t_2 e^{-i k_3} + t_3 e^{ik_2}\\
t_1 + t_2 e^{i k_3} + t_3 e^{-ik_2} & \Omega \end{array}\right),
\end{equation}
with $2\times 2$ blocks acting in the orbital space
\begin{align}
\Omega = \lambda \left( \begin{array}{cc}
0 & -i\\
i & 0 \end{array}\right), 
& & 
t_1 = t\left( \begin{array}{cc}
\frac{3}{4} & \frac{\sqrt{3}}{4}\\
\frac{\sqrt{3}}{4} & \frac{1}{4} \end{array}\right),
& & 
t_2 = t \left( \begin{array}{cc}
\frac{3}{4} & -\frac{\sqrt{3}}{4}\\
-\frac{\sqrt{3}}{4} & \frac{1}{4} \end{array}\right),
& & 
t_3 = t \left( \begin{array}{cc}
0 & 0\\
0 & 1 \end{array}\right).
\end{align}
The model has a particle hole symmetry $U_c^{\dagger}H^*(-k)U_c = H(k)$ with $U_c = \sigma_z$
($\sigma$ matrices act in sublattice space). This model is a Chern insulator with $|C|=1$ for
$|\lambda|<3/2t$. To compute the weak index, we first go to the Majorana basis, where all hoppings
becombe imaginary, by the transformation $ \left( p_x^A, p_y^A, p_x^B, p_y^B \right) 
\rightarrow \left( p_x^A, p_y^A, i p_x^B, i p_y^B \right) $. Defining again $H=iA$, at the
$\Gamma$ point we have
\begin{align}
A(\Gamma) = \left( \begin{array}{cccc}
0 & \lambda & \frac{3t}{2} & 0 \\
-\lambda & 0 & 0 & \frac{3t}{2} \\
0 & -\frac{3t}{2}& 0 & \lambda \\
-\frac{3t}{2} & 0 &-\lambda & 0 \end{array}\right) ,
\end{align}
and the Pfaffian \cite{K01} is given by 
\begin{equation}
Pf[A(\Gamma)] = \lambda^2 - 9/4t^2, 
\end{equation}
while at the M points
\begin{align}
A(M_1) = \left( \begin{array}{cccc}
0 & \lambda & 0 & \frac{\sqrt{3}t}{2} \\
-\lambda & 0 &  \frac{\sqrt{3}t}{2} & -t \\
0 & -\frac{\sqrt{3}t}{2}& 0 & \lambda \\
-\frac{\sqrt{3}t}{2} & t &-\lambda & 0 \end{array}\right),  & &
A(M_2) = \left( \begin{array}{cccc}
0 & \lambda & 0 & \frac{\sqrt{3}t}{2} \\
-\lambda & 0 &  \frac{\sqrt{3}t}{2} & t \\
0 & -\frac{\sqrt{3}t}{2}& 0 & \lambda \\
-\frac{\sqrt{3}t}{2} & -t &-\lambda & 0 \end{array}\right),  & & 
A(M_3) = \left( \begin{array}{cccc}
0 & \lambda & \frac{3t}{2} & 0 \\
-\lambda & 0 & 0 & -\frac{t}{2} \\
-\frac{3t}{2} & 0 & 0 & \lambda \\
0  & \frac{t}{2} &-\lambda & 0 \end{array}\right), \label{matrices}
\end{align}
and remarkably the Pfaffian is the same for all three points
\begin{equation}
Pf[A(M_i)] = \lambda^2 + 3/4 t^2 .
\end{equation}
According to Eq.~\eqref{weakhon}, it follows that the weak index is now trivial,
$(\nu_1,\nu_2)=(0,0)$. If a
different unit cell had been chosen, this would simply interchange the matrices in
Eq.~\eqref{matrices}, but since their Pfaffian is positive the weak index stays trivial for any unit
cell
choice. This implies that this model has no bound states at dislocations of either type. 

Another implication of a trivial weak index which is independent of the unit cell is that the number
edge states crossing at $k=\pi$ in a ribbon geometry must be even,
and must be the same for both zigzag and bearded edges. Since $|C|=1$, the number of edge states is
one and this state must thus cross at $k=0$. The number of crossings at $k=\pi$ is therefore zero. 
This is confirmed by the explicit calculation of the spectrum in a ribbon, shown in
Fig.~\ref{pband}. As can be seen, for both zigzag and bearded edges the edge state crosses at $k=0$
(contrary to the Haldane model). This is also another way of seeing that dislocations carry no zero
modes, since the glueing construction requires one crossing at $k=\pi$. 
\section{Square lattice models}
\subsection{Single-orbital square lattice model}
\begin{figure}[t]
\begin{center}
\includegraphics[width=7cm]{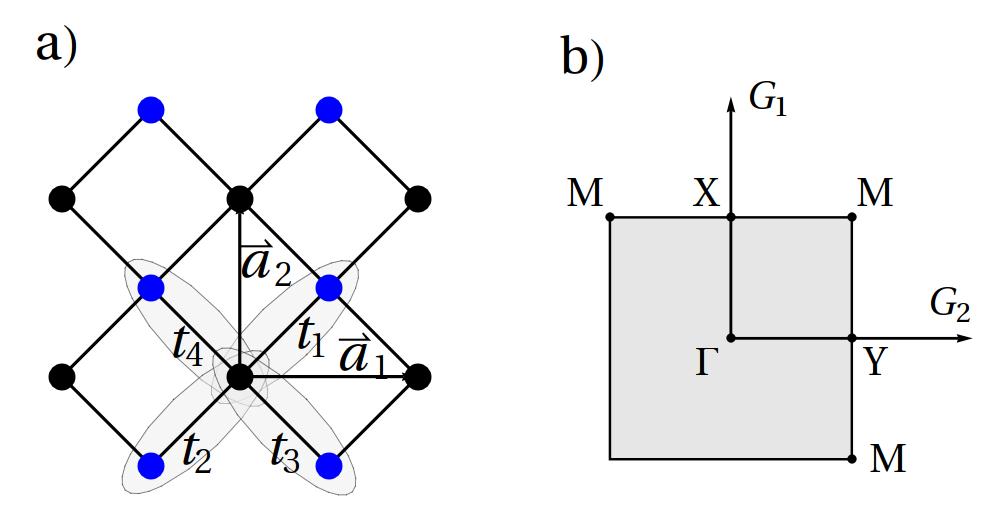}
\includegraphics[width=5cm]{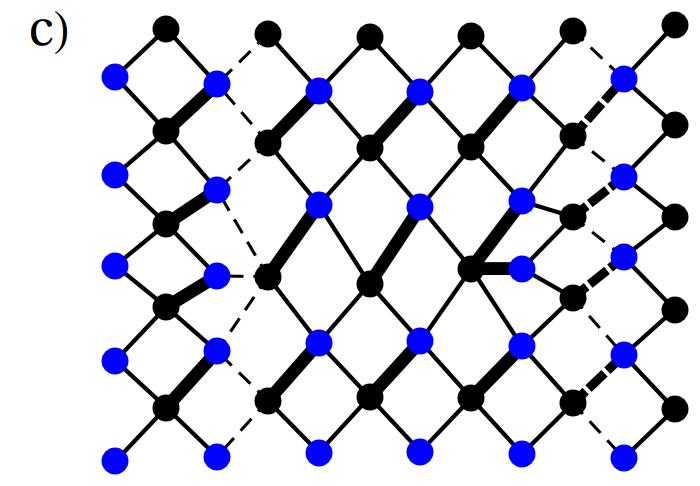}
\includegraphics[width=4.3cm]{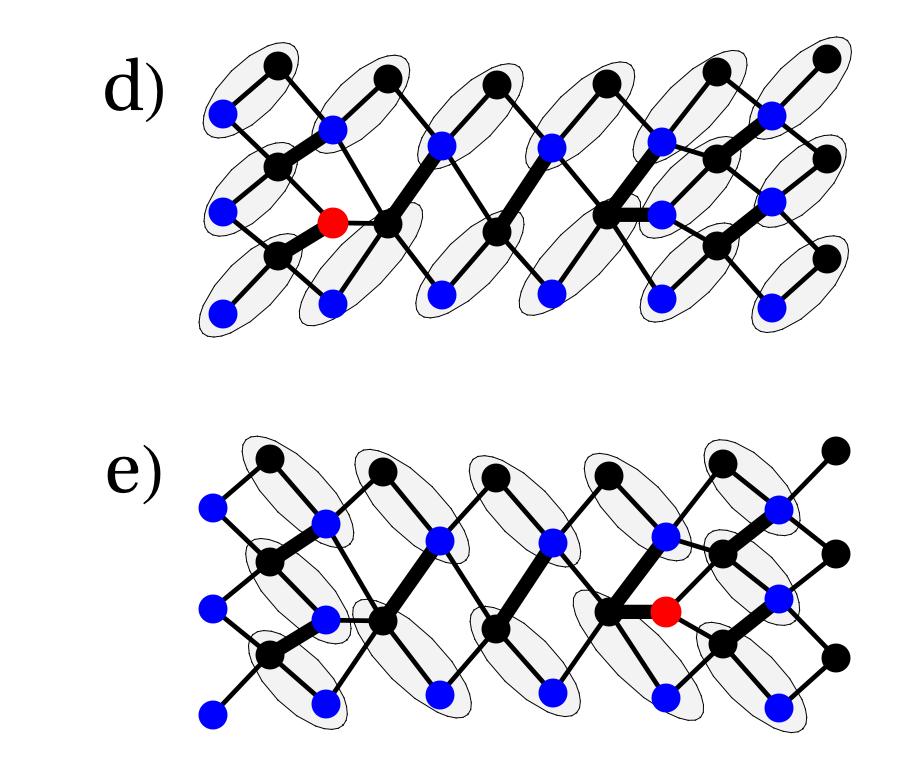}
\caption{a) Dimerized square lattice model, with the four unit cells denoted as gray ovals. b)
Brillouin Zone for the square lattice. c) A pair of dislocations for dimerization
in $t_1$ (represented as thicker lines). As in main text, the two types of dislocations are obtained
by glueing together two edges (with dashed lines) and skipping a row in the process. In this model,
the two dislocations are structurally identical and are only distinguished by the pattern
of hoppings around the core (so they are physically different). d) Unit cell
 2 tiles the second defect correctly but leaves a broken unit cell in the first (red site). e) Unit
cell 3 tiles the first defect correctly, but leaves a broken cell in the second.}
\label{squarelat} 
\end{center}
\end{figure}

We now explore the findings of our work in square lattice models. We first consider a
dimerized square lattice with two sites per unit cell and a single orbital per site,
illustrated in Fig.~\ref{squarelat}(a) (an analog of the nematic model in the honeycomb). As on the
honeycomb lattice, we define unit cell $i$ as the unit cell which encloses the bond with hopping
$t_i$. The two
types of dislocations in this model are structurally identical, but differ in the values of $t_i$
around the core. The two different types of dislocations for $\vec b = (0,1)$ are shown in
Fig.~\ref{squarelat}(c)-(e) for dimerization in $t_1$, and unit cell choices 2 and 3. 

We now evaluate the weak index of this model. In unit cell 2, the bulk Hamiltonian takes the form
\begin{align}
H(\vec k) = \sigma_x\left(t_1\cos (k_1+k_2) +t_2  + t_3 \cos k_1 + t_4 \cos k_2\right) + 
\sigma_y \left(t_1 \sin (k_1+k_2) + t_3 \sin k_1 + t_4 \sin k_2\right).  
\end{align}
which has a $\mathcal{C}$-symmetry with $U_c = \sigma_z$. Different unit cells effectively
interchange the different $t_i$. The values of $f_x$ for different unit cells are shown in
Table~\ref{table2} along with the values of the weak indices computed from Eq.~\eqref{wisq}. Note
that if
$t_1=t_2=t_3=t_4$ then $f_x(X)=f_x(Y)=f_x(M)=0$ and the weak index is not defined. Small changes in
$t_i$ will turn the model into either a weak TI or a gapless system, with well-defined $\vec N$.

\begin{figure}[t]
\begin{center}
\includegraphics[width=8.9cm]{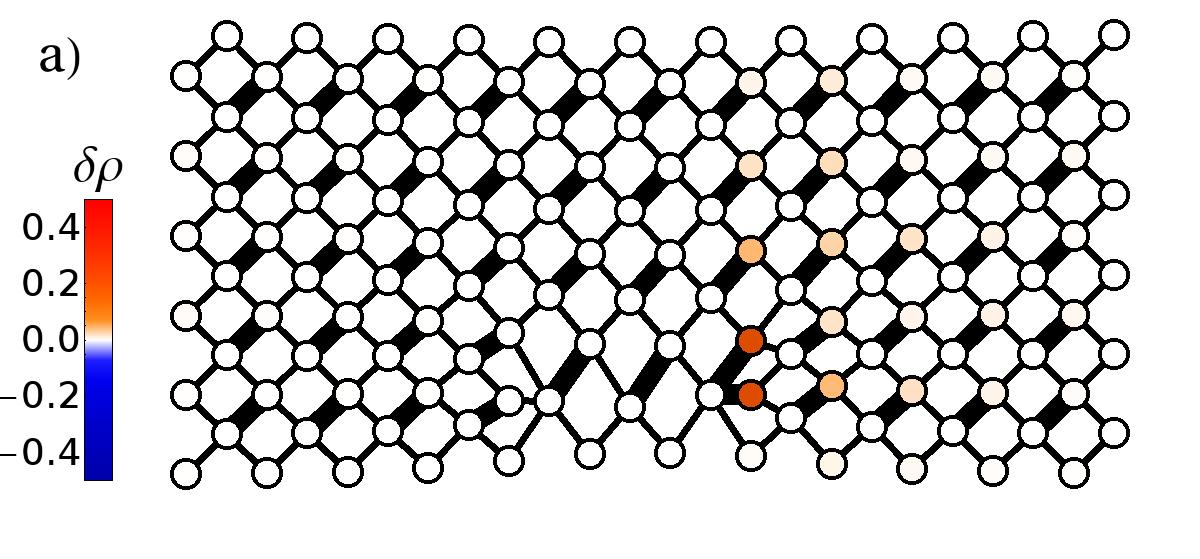}
\includegraphics[width=8.9cm]{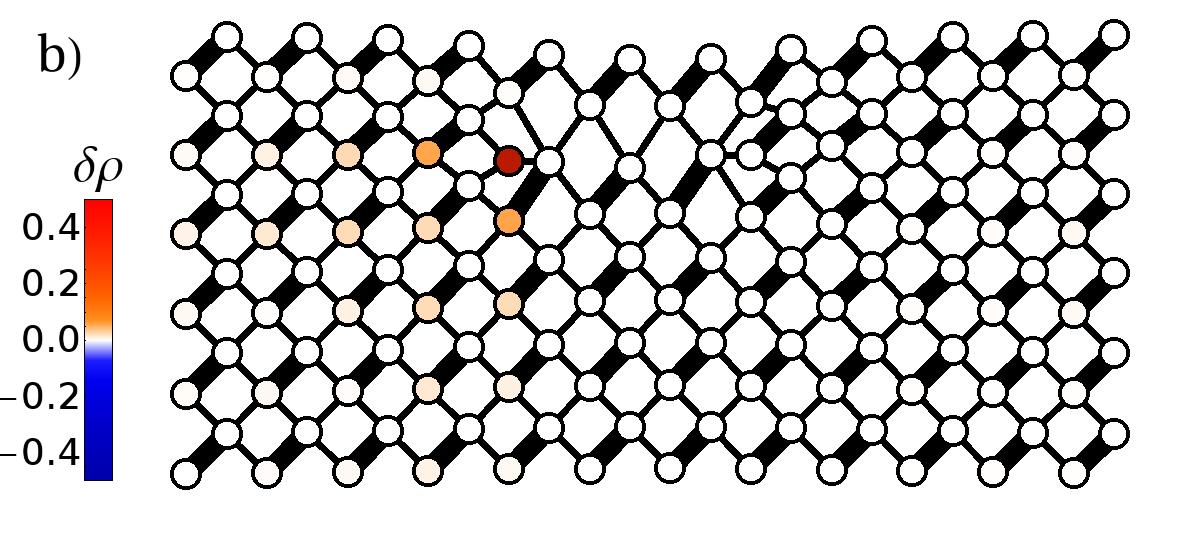}
\includegraphics[width=8.9cm]{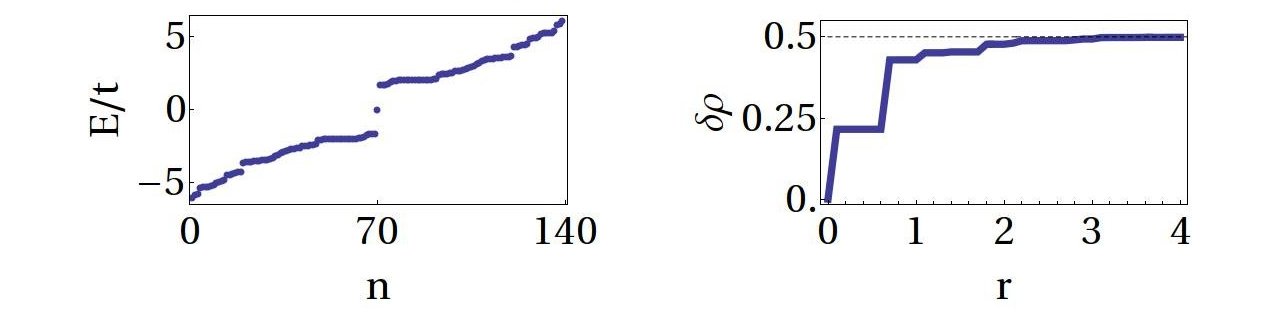}
\includegraphics[width=8.9cm]{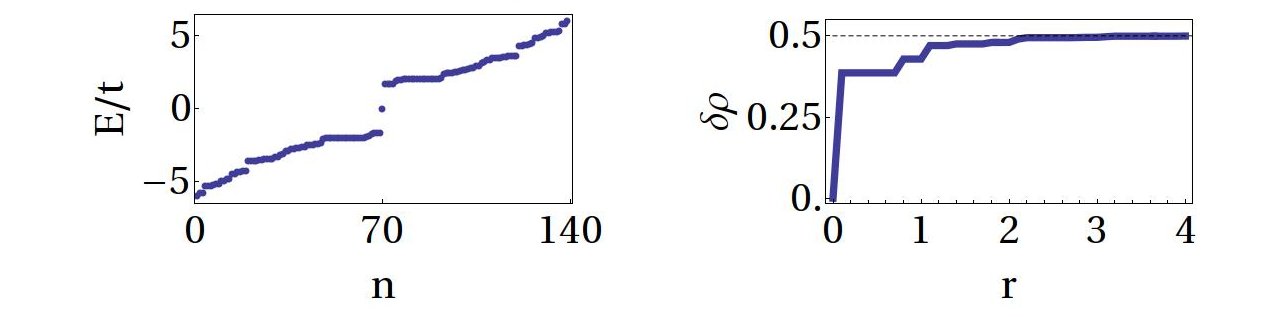}
\caption{a) Top: Charge distribution at half-filling (zero mode is occupied) for a dislocation
dipole and dimerization in $t_1=3t$ and $t_{2,3,4}=t$, computed from the tight-binding model
represented in the figure with periodic boundary conditions. Bottom left: Energy spectrum of the
system, with a single zero mode within the gap. Bottom right: Integrated charge within a radius $r$
from the core of the dislocation carrying the zero mode. Charge converges to 1/2 within a few
lattice spacings. b) The same for dimerization in $t_2= 3t$ and $t_{1,3,4}=t$
(dislocation is shifted vertically for clarity).}\label{squarenumerics} 
\end{center}
\end{figure}

We now test these expressions against the numerical solution of the tight-binding model for
dislocations. A dislocation pair with Burgers vector $\vec b= (0,1)$ is shown in
Fig.~\ref{squarelat}(c). For unit cell 2, the first dislocation has a broken unit cell while the
second is
correctly tiled. Consider first dimerization in $t_1$, which from Eq.~\eqref{wisq} gives $\vec N =
(1,1)\pi$. The second dislocation must therefore have a zero mode since $\vec N \cdot \vec b =\pi$,
but not the first one, because a broken cell adds an extra zero mode to the count. Consider now
dimerizing in $t_2$, which from Eq.~\ref{wisq} gives $\vec N = (0,0)\pi$. Now $\vec N \cdot \vec b
=0$ and the second dislocation must be trivial, while the first binds a zero mode. These results
are confirmed by the numerical solution as shown in Fig.~\ref{squarenumerics} for these two
dimerizations. This example illustrates a particular case, but we have checked explicitly that our
conclusions hold for every unit cell and dimerization choices. 

\begin{center}
\begin{table}
\begin{tabular}{|c|c|c|c|c|}\hline
\multicolumn{5}{|c|}{$f_x$}\\ \hline
\; & Unit cell 1 & Unit cell 2 & Unit cell 3 & Unit cell 4\\ \hline
$f_x(\Gamma)=$ & $t_1+t_2+t_3+t_4$ & $t_1+t_2+t_3+t_4$ & $t_1+t_2+t_3+t_4$& $t_1+t_2+t_3+t_4$\\
\hline
$f_x(X)=$ & $t_1-t_2+t_3-t_4$ & $-t_1+t_2-t_3+t_4$ & $t_1-t_2+t_3-t_4$& $-t_1+t_2-t_3+t_4$\\ \hline
$f_x(Y)=$ & $t_1-t_2-t_3+t_4$ & $-t_1+t_2+t_3-t_4$ & $-t_1+t_2+t_3+t_4$& $t_1-t_2-t_3+t_4$ \\ \hline
$f_x(M)=$ & $t_1+t_2-t_3-t_4$ & $t_1+t_2-t_3-t_4$ & $-t_1-t_2+t_3-t_4$& $-t_1-t_2+t_3+t_4$ \\\hline
\end{tabular}
\caption{Values of $f_x(k)$ at the PHIM for different unit cells in the square lattice
model.}\label{table2}
\end{table}
\end{center}
\begin{center}
\begin{table}
\begin{tabular}{|c|c|c|c|c|}\hline
\multicolumn{5}{|c|}{($\nu_1$,$\nu_2$)}\\ \hline
Model & Unit cell 1 & Unit cell 2 & Unit cell 3 & Unit cell 4\\ \hline
 $t_1$-dimerized & (0,0) & (1,1) & (1,0)& (0,1)\\ \hline
 $t_2$-dimerized & (1,1) & (0,0) & (0,1)& (1,0)\\ \hline
 $t_3$-dimerized & (1,0) & (0,1) & (0,0)& (1,1) \\\hline
 $t_4$-dimerized & (0,1) & (1,0) & (1,1)& (0,0) \\\hline
\end{tabular}
\caption{Weak indices for the square lattice model as computed from Eq.~\eqref{wisq} for different
dimerizations.}
\end{table}
\end{center}

\subsection{Two-orbital square lattice model}
The two-orbital model on the square lattice which we discuss in the following is obtained by
restricting the Bernevig-Hughes-Zhang (BHZ) model to only one spin component. This model then
realizes a Chern insulator and is given by \cite{CH94,BHZ06}
\begin{align}
H(\vec k) =& \tau_x \sin k_1 + \tau_y \sin k_2 + \tau_z [M-2B(2-\cos k_1 - \cos k_2)].
\label{CH}
\end{align}
Here, the Pauli matrices $\vec \tau$ act in orbital space. This model has an on-site particle-hole
symmetry with $U_c = \tau_x$. Because there are two orbitals per site, $\eta=0$ and the removal of a
site does not generate a zero mode. The model Eq.~\eqref{CH} on the square lattice therefore has one
type of
dislocation only. To determine if it has a zero mode, we bring the Hamiltonian to the Majorana
basis with a unitary transformation $H \rightarrow UHU^{\dagger}$ given by
\begin{equation}
 U= \left( \begin{array}{cc}
1 & 1 \\
-i & i
\end{array}\right),
\end{equation}
where the Hamiltonian is now
\begin{align}
H(\vec k) =& \tau_z \sin k_1a - \tau_x \sin k_2 - \tau_y [M-2B(2-\cos k_1 - \cos k_2)].
\end{align}
Again at the PHIM $H=iA$, and the Pfaffian is the coeficient in front of $-\tau_y$, which at
the different PHIM is 
\begin{align}
Pf[A(\Gamma)]= M, && Pf[A(X)]= M-4B, && Pf[A(Y)]= M - 4B, && Pf[A(M)] = M - 8B,
\end{align}
and the weak index is 
\begin{align}
(-1)^{\nu_1} = \text{sign}[Pf[A(Y)]]\; \text{sign}[Pf[A(M)]] =
\text{sign}(M-4B)\text{sign}(M-8B), \\
(-1)^{\nu_2} = \text{sign}[Pf[A(X)]]\; \text{sign}[Pf[A(M)]] =
\text{sign}(M-4B)\text{sign}(M-8B).
\label{wipff2}
\end{align}
which is $(\nu_1,\nu_2)=(1,1)$ when $4<M/B<8$ and zero otherwise. The model has a zero mode in
dislocations only in the non-trivial regime (named the M-phase), as obtained in Ref.~\cite{JMS12}.
Finally, our analysis can also be extended to more complicated square lattice
models with non-trivial weak indices such as the one described in Ref.~\cite{FIH13}.
\section{$\mathcal{C}$-symmetric QSH insulators}
Our work also sheds light on the dislocation modes of a particular class of quantum spin Hall
insulators, namely those with a particle-hole symmetry in addition to time reversal, which belong to
class DIII. Only with the extra $\mathcal{C}$-symmetry, a weak index is defined in this case due to
the
$\mathds{Z}_2$ invariant of this class in 1D. The zero modes bound to dislocations in this class
are now single Kramers pairs. In the case of spin-1/2 fermions with time-reversal symmetry, the
criteria $(-1)^{n_{\rm orb}}=-1$ for a non-trivial vacancy state still holds but $n_{\rm orb}$ is
now the number of orbitals {\em per spin}.

There are two popular models used to describe QSH insulators, the Bernevig-Hughes-Zhang (BHZ) and
the Kane-Mele (KM) model. In the limit of zero Rashba spin-orbit coupling, both models can be seen
as two copies of a Chern insulator with opposite Chern number for the two spins. The Kane-Mele (KM)
model \cite{KM05} is given by two spin copies of the Haldane model, and, as discussed in the text,
has a particle-hole symmetry given by $U_c=\sigma_z$. The BHZ model \cite{BHZ06} is given by two
copies of Eq.~\eqref{CH} with particle-hole symmetry given by $U_c=\tau_x$. 

From the analysis of the spinless models, we know that the BHZ model has one type of dislocation,
with a zero-energy Kramers pair only in the M-phase \cite{JMS12}, and a weak index that is
independent of the unit cell choice. On the other
hand, the KM model has only one phase with zero modes at the 6-8 dislocations but not at the 5-7
ones (as described in the main text). This two models are therefore very different regarding their
response to dislocations. 

Finally, it should be noted that Rashba spin-orbit coupling breaks this $\mathcal{C}$-symmetry in
both models, and thus moves the in-gap state away from zero energy. In the KM model, the Rashba term
takes the form in real space 
\begin{equation}
H_R = i \lambda_R \sum_{<ij>} c^{\dagger}_i (\vec s \times \vec
d_{i,j})  c_j +c.c. ,
\end{equation}
with $d_{i,j}$ the nearest neighbour vectors and $\vec s = (s_x,s_y)$ the spin matrices. For unit
cell 1, the fourier transform of $H_R$ is
\begin{equation}
H_R =  \left(\begin{array}{cc}
       0 & i \lambda_R\left[ s_x - \frac{1}{2}(s_x - \sqrt{3}s_y) e^{-i k_3} -
\frac{1}{2}(s_x + \sqrt{3}s_y) e^{i k_2} \right] \\
-i\lambda_R \left[ s_x - \frac{1}{2}(s_x - \sqrt{3}s_y) e^{i k_3} -
\frac{1}{2}(s_x + \sqrt{3}s_y) e^{-i k_2} \right] & 0
      \end{array}\right),
\end{equation}
which does not satisfy Eq.~\eqref{ph} with $U_c=\sigma_z$. Similarly, the Rashba term for the BHZ
model \cite{JMS12} is 
\begin{equation}
H_R = i \lambda_R \sum_{<ij>} c^{\dagger}_i [(1+\tau_z)\vec s \times \vec
d_{i,j} ] c_j,
\end{equation}
which in momentum space is
\begin{equation}
H_R = \lambda_R (\sigma_x\sin k_2 - \sigma_y \sin k_1)(1+\tau_z),
\end{equation}
and again Eq.~\eqref{ph} with $U_c=\tau_x$ is not satisfied.
\section{120$^{\circ}$ disclinations in the Haldane model}
\begin{figure}[t]
\begin{center}
\includegraphics[width=5cm]{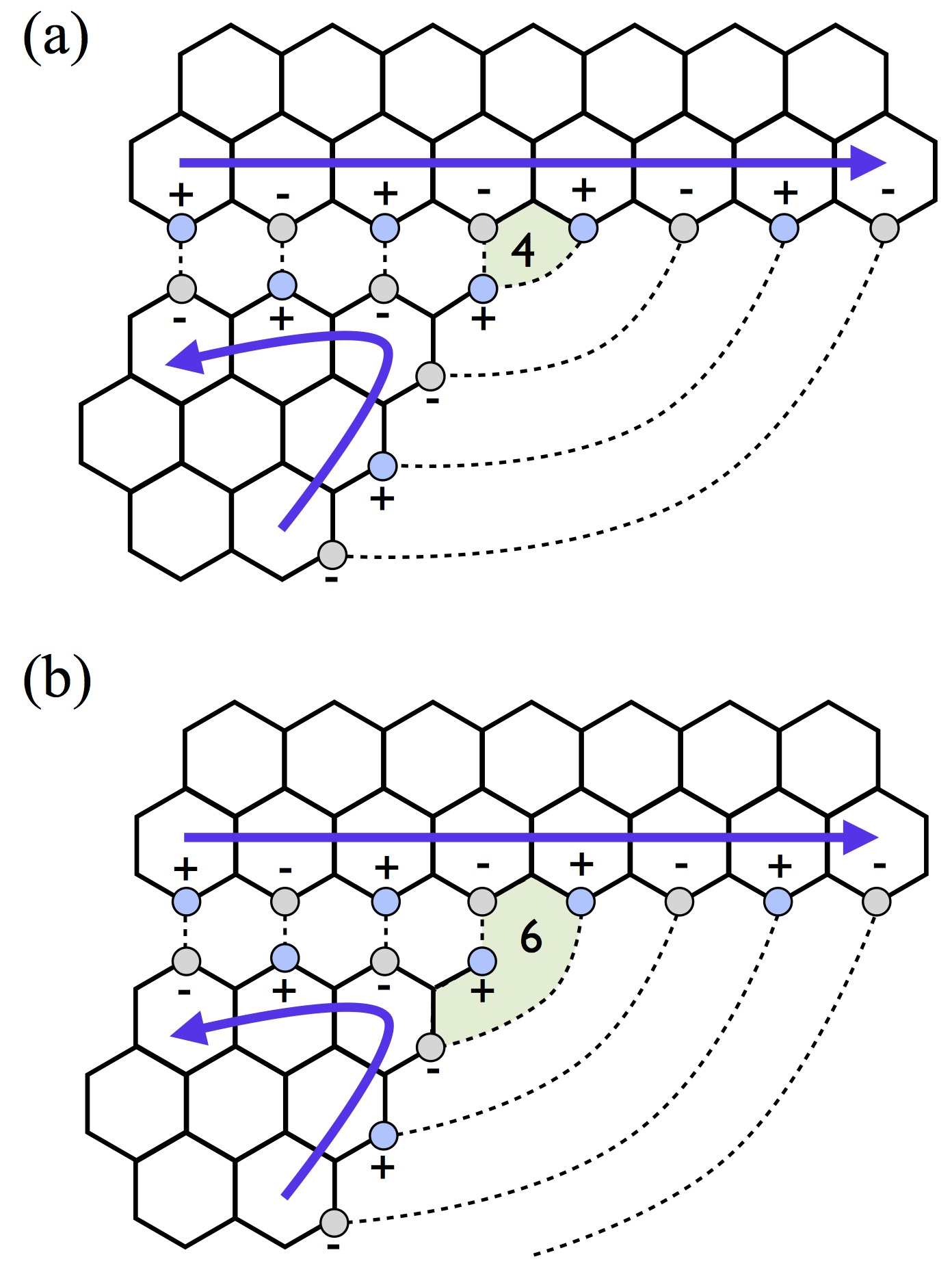}
\includegraphics[width=4cm]{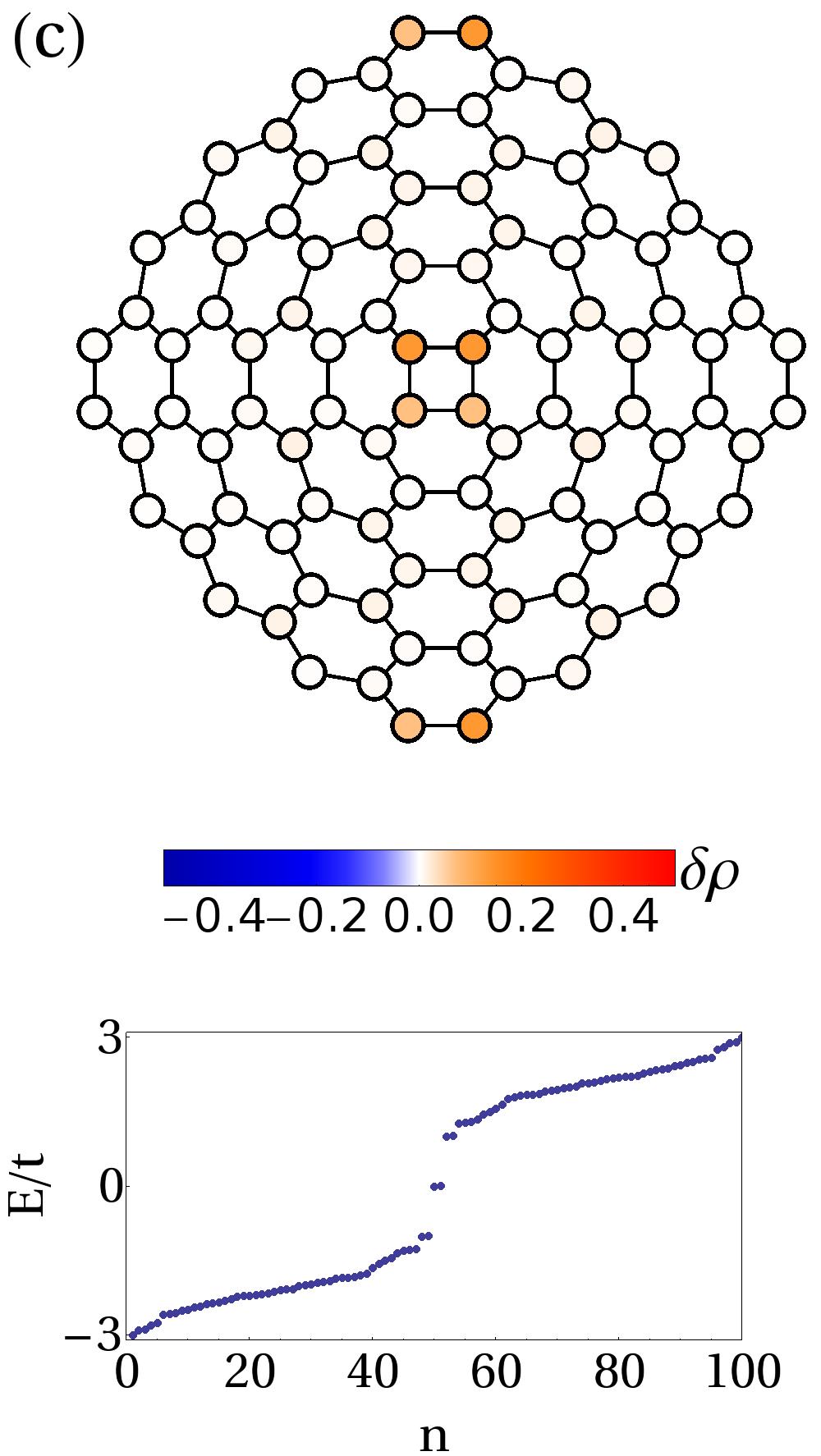}
\includegraphics[width=4cm]{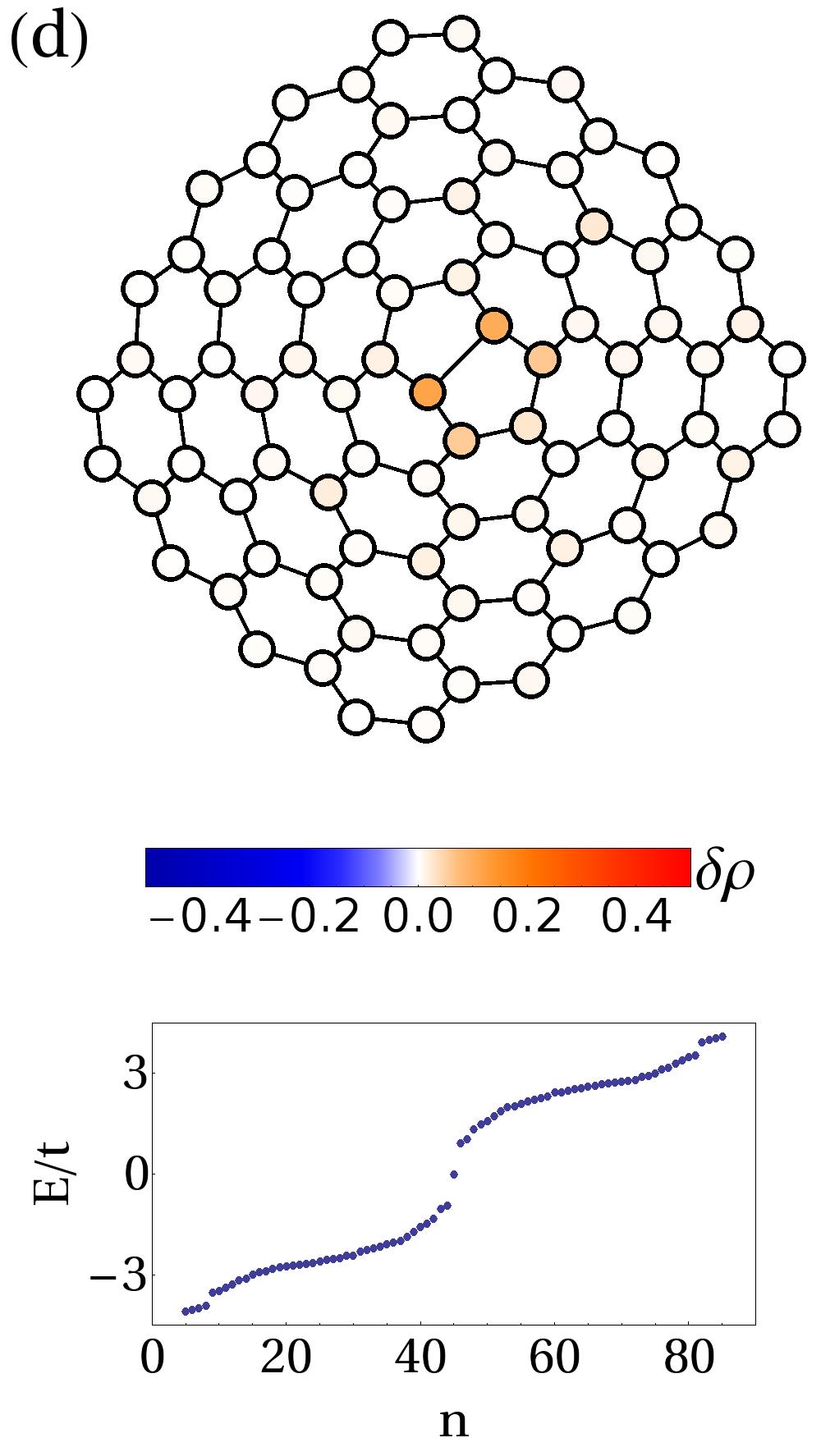}
\includegraphics[width=4cm]{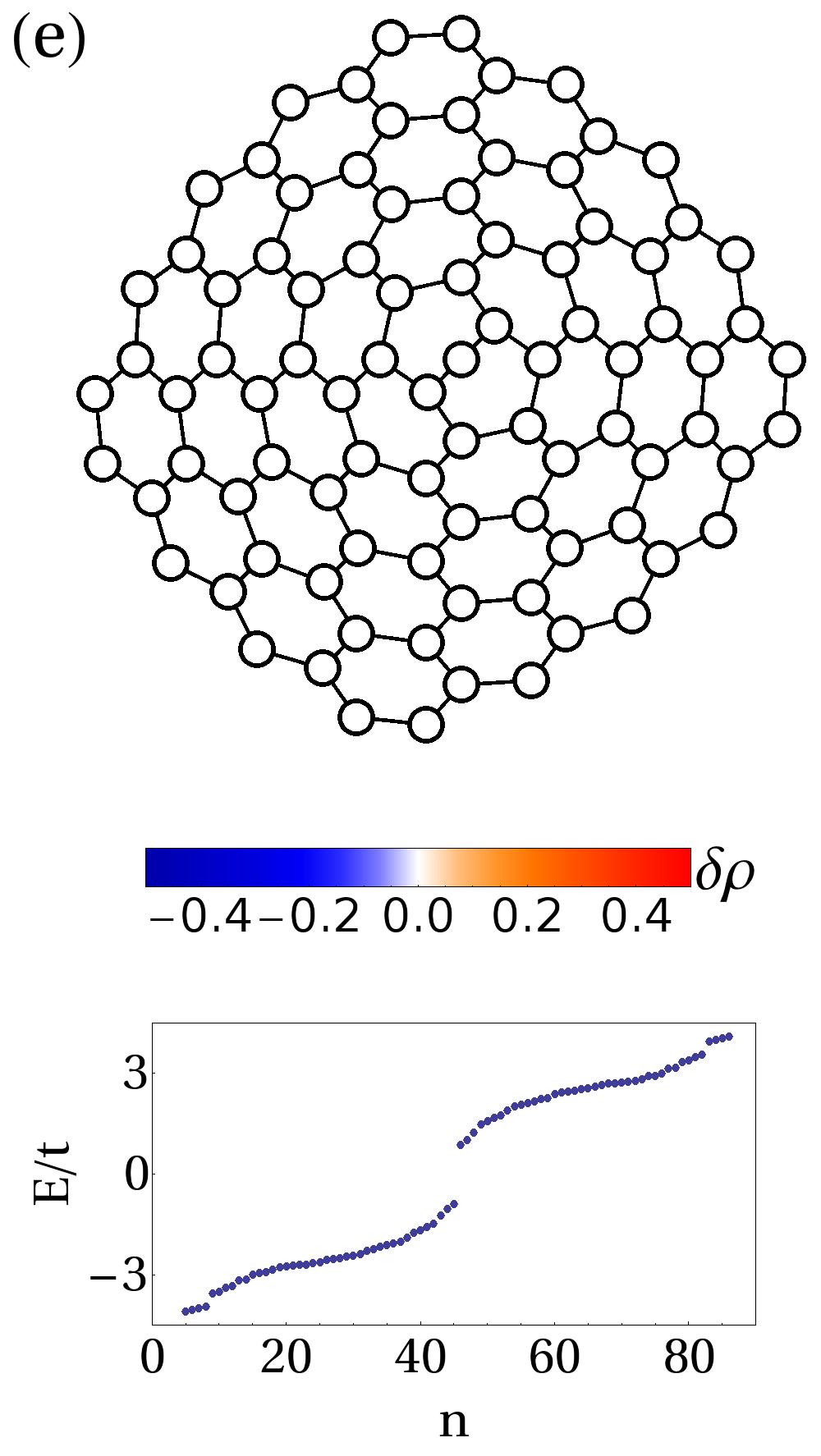}
\caption{a,b)Volterra process for the 4, and 6-6 disclinations built from glueing zigzag
edges. The 4 disclination has a phase mismatch of $\pi$, and a fractional charge 1/2,
while the 6-6 does not (compare with ref. \onlinecite{RL13}). c-e) Charge distribution
and spectrum for the tight binding models for the 4, 5-5 and 6-6 disclinations (see
text for details). The 4 and 5-5 have a zero mode with charge 1/2, but the 6-6 does
not. }\label{disc}
\end{center}
\end{figure}

In this section, we briefly discuss how a non-trivial 0D index can also affect zero modes in
disclinations. For the Haldane model, 120$^\circ$ disclinations preserve the particle-hole symmetry
and we focus therefore on these defects. There are three different 120$^\circ$ disclinations: the
square (4), the pentagon-pentagon (5-5) and the hexagon-hexagon (6-6) disclinations. The (5-5)
disclination is obtained from the (6-6) by the removal of the central site. Following the edge-state
picture given in Ref.~\cite{RL13}, one finds that the square disclination has a zero mode while the
(6-6) does not, see Fig.~\ref{disc}(a) and (b). Because  $\eta_A=\eta_B=1$ for the Haldane model, it
follows that the (5-5) disclination has a bound state as well. 

These expectations were confirmed numerically in a tight binding model as in the previous sections.
The method to build a lattice with periodic boundary conditions and disclinations was described in
ref. \cite{TH13}, and in general requires several disclinations to make the lattice close. The
lattices used in this work are shown in fig \ref{disc} and contain a 120$^\circ$ disclination at the
center, another 120$^\circ$ one shared by the upper and lower corners, and two indepedent
240$^\circ$ ones (which are always trivial) in the left and right corners. 

Fig.~\ref{disc}(c) shows a lattice with two square disclinations. Two zero energy states are seen
in the spectrum, whose wavefunctions are localized at the two defects. When these states are filled
the plots display an accumulated charge at each defect, which integrates to 1/2 for a disk around
the defect. It should be noted that the phases of the NNN hoppings across the square are
frustrated, and a particular choice of phase always breaks the $C_4$ rotational symmetry around the
square, as can be seen in the charge distribution. If these two hoppings are set to zero the charge
distribution becomes symmetric. Fig.~\ref{disc}(d) shows a lattice with a (5-5) disclination at
the center (and another 6-6 at the upper-lower corner). Now there is only one state bound to the
5-5, again with fractional charge 1/2 (note the anisotropy again due to the frustration of the
phase of the pentagon link). Finally Fig.~\ref{disc}(e) shows a lattice with two (6-6) defects and
no zero modes at all. 

\end{document}